\documentclass[aps,reprint,amsmath,amssymb,graphicx,onecolumn]{revtex4-2}
\usepackage[utf8]{inputenc}
\usepackage{natbib}
\usepackage{color}
\usepackage{xcolor}
\usepackage{graphicx}
\usepackage{mdframed}
\usepackage{physics}
\usepackage{bm}
\patchcmd{\section}{\centering}{\raggedright}{}{}
\patchcmd{\subsection}{\centering}{\raggedright}{}{}
\usepackage[percent]{overpic}
\usepackage{amsmath,amssymb}
\usepackage{bbold}

\begin{document}
\preprint{APS/123-QED}

\title{Disentangling the critical signatures of neural activity}

\author{Benedetta Mariani$^{1,2}$, Giorgio Nicoletti$^{1}$, Marta Bisio$^{2,3}$, Marta Maschietto$^{3}$, Stefano Vassanelli$^{2,3}$, Samir Suweis$^{1,2}$\\}
\address{$^1$Department of Physics and Astronomy ``Galileo Galilei'' \\
University of Padova, Padova, Italy \\
$^2$Padova Neuroscience Center \\
University of Padova, Padova, Italy \\
$^3$Department of Biomedical Science \\
University of Padova, Padova, Italy \\
}

\begin{abstract}
The critical brain hypothesis has emerged as an attractive framework to understand neuronal activity, but it is still widely debated. In this work, we analyze data from a multi-electrodes array in the rat's cortex and we find that power-law neuronal avalanches satisfying the crackling-noise relation coexist with spatial correlations that display typical features of critical systems. In order to shed a light on the underlying mechanisms at the origin of these signatures of criticality, we introduce a paradigmatic framework with a common stochastic modulation and pairwise linear interactions inferred from our data. We show that in such models power-law avalanches that satisfy the crackling-noise relation emerge as a consequence of the extrinsic modulation, whereas scale-free correlations are solely determined by internal interactions. Moreover, this disentangling is fully captured by the mutual information in the system. Finally, we show that analogous power-law avalanches are found in more realistic models of neural activity as well, suggesting that extrinsic modulation might be a broad mechanism for their generation.
\end{abstract}

\flushbottom
\maketitle
\thispagestyle{empty}
\section*{Introduction}
The critical brain hypothesis has been much investigated since scale-free neuronal avalanches were found in 2003 by Beggs and Plenz \cite{beggs2003}. By analyzing Local Field Potentials (LFPs) from cortical slices and cultures on chips, their seminal work showed that neural activity occurred in cascades, named neuronal avalanches. Remarkably, they found that both the sizes and the lifetimes of these avalanches were power-law distributed, with exponents surprisingly close to the ones of a critical mean-field branching process. Since then, such power-laws have been repeatedly observed in experiments \cite{Petermann2009, Yu2011, Hahn2010, Gireesh2008, Mazzoni2007, Pasquale2008}, giving rise to the idea that the brain might be poised near the critical point of a phase transition\cite{de2006self, kinouchi2006optimal, Shew2013, hesse2014self, hidalgo2014information, tkavcik2015thermodynamics, rocha2018homeostatic, Munoz2018}. Yet, this hypothesis is still widely debated, and reconciling the different views and experimental results remains pressing.

On the one hand, many subsequent works showed that the presence of power-law avalanches is not a sufficient condition for criticality, as they might emerge from different mechanisms \cite{Touboul2010, Touboul2017, Martinello2017, priesemann2018can, faqeeh2019emergence}. {A perhaps stronger} test for criticality is whether the avalanche exponents satisfy the crackling-noise relation, a relation that connects the avalanche exponents to the scaling of the average avalanche size $\ev{s}$ with its duration $T$ \cite{Sethna, Friedman2012, diSanto2017}, defined by the exponent $\ev{s(T)} \sim T^\delta$. In particular, recent findings \cite{Fontenele2019, Buendia2021, Carvalho2021} suggest that, while the avalanche exponents found in different experimental settings do vary, they all lay along the scaling line defined by the crackling-noise relation with a seemingly universal exponent $\delta \approx 1.28$. Nevertheless, it was recently suggested that this relation can be fulfilled {in different settings \cite{Scarpetta} and} even in models of independent spiking units, for a range of choices of the power-law fitting method \cite{Destexhe2020}.

On the other hand, the debate about the ``nature'' of the transition is very much open, and thus its hypothetical universality class is poorly understood. In particular, recent works have proposed that the observed transition might be related to a synchronous-asynchronous one \cite{diSantosincr, Dallaporta2019, Poil2012, Buendia2021} or a disorder-induced transition \cite{Ponce-Alvarez2018}, contrarily to the original hypothesis of a simpler critical branching process.

Arguably, a more fundamental signature of criticality is the presence of power-law correlations in space \cite{binney1992, henkel2009}. In particular, a key feature of both equilibrium and non-equilibrium systems is a correlation length that, in the thermodynamic limit, diverges at criticality. In finite systems, such scale-free correlations manifest themselves in a correlation length that scales with the system size. Yet, the study of these correlations has usually been applied at coarser scales, such as in whole-brain data \cite{Haimovici2013, Ponce-Alvarez2018}, and only recently in specific cortical areas \cite{Ribeiro2020}. When such spatial information is not available, phenomenological renormalization group procedures based on the correlation structure between the underlying degrees of freedom have been recently proposed, although with some limitations in their interpretability \cite{meshulam2019coarse, Nicoletti2020}.

All in all, these results still lack a unifying framework. In this work, we draw a possible path to reconcile these different signatures of criticality. First, we analyze spatially extended LFPs from the rat's somatosensory barrel cortex. The multi-electrodes array records LFPs from a whisker-related barrel column that spans vertically the six layers of the barrel cortex. We find both scale-free avalanches that satisfy the crackling-noise relation and signatures of criticality beyond avalanches in the form of spatial correlations with a characteristic length that grows linearly with the system size. Then, we introduce an archetypal framework in which the origin of these different signatures can be disentangled, considering both a paradigmatic model that can be tackled analytically and a more biophysically sound model. In doing so, we shed a light on the underlying mechanisms from which such properties of neural activity emerge. In particular, our work suggests that, whereas avalanches may emerge from an external stochastic modulation that affects all degrees of freedom in the same way, interactions between neural populations are the fundamental biological mechanism that gives rise to seemingly scale-free correlations.

\section*{Results}

\begin{figure*}[t]
    \centering
    \includegraphics[width=\textwidth]{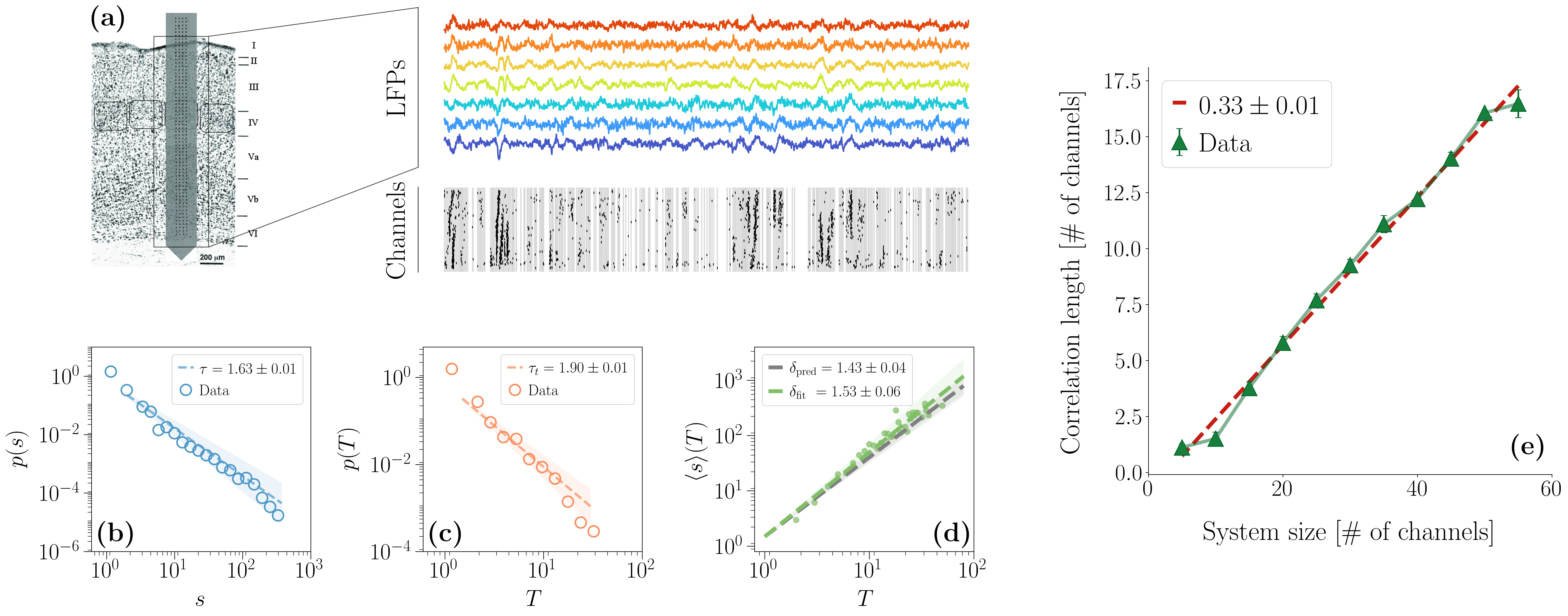}
    \caption{(a) Left: scheme of the array used to obtain the LFPs data from all the cortical layers of the barrel cortex (adapted from \cite{Zhang6365}); right an example of the LFPs signals for different layers and the corresponding discretization. An array of 256 channels organized in a $64 \times 4$ matrix is inserted in a barrel column and the signals from the cortical layers are collected by $55 \times 4$ electrodes. (b-d) Avalanche statistics obtained from the analysis of LFPs data in a rat. Both the distribution of the avalanches (b) sizes and (c) durations are power-laws, and (d) the crackling-noise relation is satisfied. (e) Scaling of the correlation length with the system size in LFPs data, averaging over four different rats. The error bars are shown as $5$ standard deviations from the mean for visual ease. The correlation length scales linearly with the system size with no plateau in sight, a hallmark of criticality.}
    \label{fig:LFPData}
\end{figure*}

\subsection*{Scale-free avalanches and correlations in LFPs}
In this work, we study LFPs activity in the primary somatosensory cortex of four rats using state-of-the-art spatially extended multi-electrodes arrays. The cortical activity is recorded through a 256-channels array organized in a $64$ rows $\times$ $4$ columns matrix with an inter-electrode distance of 32 $\mu m$ (see Methods and Figure \ref{fig:LFPData}a). The avalanche statistics is analyzed in LFPs data following standardized pipelines for their detection (see Methods) \cite{Gireesh2008, Petermann2009, Shriki}, and we report a full statistical analysis of our dataset in \cite{Mariani2021b}. The distribution $p(s)$ of the avalanches sizes and of the avalanche duration $p(T)$ are computed and fitted using a corrected maximum likelihood method \cite{Gerlach2019, Mariani2021b}. We find that both are statistically compatible with the expected power-laws $P(s)\sim s^{-\tau}$ and $P(T)\sim T^{-\tau_t}$ as we show in Figure \ref{fig:LFPData}b-c. Averaging over four rats, we find an inter-rat variability with average exponents $\ev{\tau}=1.75 \pm 0.1$ and $\ev{\tau_t} =2.1 \pm 0.3$ (see Supplementary Information). In Figure \ref{fig:LFPData}d we show that the crackling-noise relation \cite{Sethna} holds, by comparing $\delta_\mathrm{pred} = \frac{\tau_t-1}{\tau-1}$ with the exponent obtained by fitting the average avalanche sizes as a function of their duration, i.e., $\ev{s(T)} \sim T^{\delta_\mathrm{fit}}$. Averaging over each of our rats, we find $\ev{\delta_{\mathrm{pred}}}=1.47\pm0.18$ and $\ev{\delta_\mathrm{fit}} = 1.46 \pm 0.14$. Further details on the fitting procedure can be found in the Methods and in \cite{Mariani2021b}, where we also use higher frequency data (MUAs) from the same experimental condition, but with a much less dense array. Although MUAs are not suitable for the study of the correlation length, they reproduce more closely the avalanche exponents and in particular the scaling exponent $\delta \approx 1.28$ \cite{Fontenele2019, Buendia2021, Carvalho2021}. On the other hand, LFPs are known to be strongly affected by finite-size effects \cite{beggs2003}, {as we show in \cite{Mariani2021b} by performing finite-size scaling on LFPs from the same set of experiments}.

Then, thanks to the extended spatial resolution of the multi-electrodes array, we focus on the spatial correlations of the fluctuations of the measured LFP activity and whether they display any signature of criticality \cite{Fraiman2012, Haimovici2013}. We study the scaling of the correlation length $\xi$ as a function of the system sizes $L$ by selecting different portions of the array \cite{Cavagna2010, martin2020} (see Methods) and we find that $\xi$ scales linearly with $L$. This result can be interpreted as a signature of the presence of underlying long-range correlations that scale with the system size. In fact, through simulations on control models displaying a critical point, subsampling, as we do, has been shown to be practically equivalent to considering systems of different sizes \cite{martin2020}. This behavior matches exactly what would happen at a critical point, where the correlation length diverges in the thermodynamic limit and thus grows with the size of a finite system. Hence, we find that the measured neural activity in the barrel cortex at rest displays two different signatures of criticality - power-law avalanches and scale-free spatial correlations.

\subsection*{A disentangled model for extrinsic activity and internal couplings}
In order to try and unfold the underlying processes from which these collective properties emerge, we assume that neural activity may be decomposed in two parts \cite{priesemann2018can, ferrari2018separating}: (i) the \emph{intrinsic activity}, which is the activity driven by interactions between neurons or populations of neurons - in our case, the propagation dynamics across the multi-layer network of the interconnected neurons along the barrel; (ii) the \emph{extrinsic activity}, which instead corresponds to the activity modulated by an external or global unit - in our case, the external inputs triggering or modulating the propagation (e.g. synaptic current injection from the thalamic inputs). {Taking into account \emph{extrinsic activity} becomes particularly important when neural activity is not analyzed in an isolated context - e.g., from neural slices - but rather directly from a portion of the animal brain, as we do here.}

To this end, we introduce a paradigmatic model of $N$ continuous {real} variables $(v_1, \dots, v_N)$, denoting the activity of $N$ units (e.g. neurons or, as in our case, distinct populations of neurons as measured by our LFPs). Intrinsic activity corresponds to pairwise interactions among these units, whereas extrinsic activity is modeled through a common external input that affects all the units in the same way. To fix the ideas, we first consider the simple case of a multivariate Ornstein-Uhlenbeck (mOU) process, in which the external input corresponds to a common modulation of the noise strength. Although not realistic from a biophysical point of view, this model is simple enough to be treated analytically and to provide a clear physical interpretation, while being complex enough to display non-trivial behaviors \cite{nozari2020brain}. Moreover, mOU processes have been already considered in the literature in the context of fMRI signals \cite{Saggio2016, Gilson2016, Gilson2019, Arbabyazd2021}. As we will see, the results we obtain for extrinsic activity are qualitatively unchanged even in more biophysically-sound models. Therefore, we first consider
\begin{equation}
\label{eqn:OUprocess_vi}
     \frac{dv_i(t)}{dt} =- \sum_{j} A_{ij}v_j(t) + \sqrt{\mathcal{D}(t)} \eta_i(t),
\end{equation}
where $\eta_i(t)$ are standard white noises{, $A$ is a $N\times N$ symmetric matrix} and {$\mathcal{D}(t)>0$} corresponds to a noise strength modulation from an external input shared among all the units. We also write {$A_{ij} = \frac{1}{\gamma_i} - \mathcal{W}_{ij}$}, where $\gamma_i$ is the characteristic time of the $i$-th unit, and $\mathcal{W}$ is the matrix of the effective synaptic strengths, whose diagonal entries are set to zero. In order to derive analytical results, following \cite{Touboul2017}, we define the noise modulation $\mathcal{D}(t)$ as
\begin{equation}
\label{eqn:OU_Dt}
    \mathcal{D}(t) = \begin{cases}
                     \mathcal{D}^* & \text{if} \quad D(t) \le \mathcal{D}^* \\
                     D(t) & \text{if} \quad D(t) > \mathcal{D}^*
                     \end{cases}
\end{equation}
where $D(t)$ is itself an OU process $\dot{D}(t) = -D(t)/\gamma_D + \sqrt{\theta} \eta_D(t)$ {and $\mathcal{D}^*>0$ is a properly chosen threshold}.  Therefore, the noise modulation $\mathcal{D}(t)$ {is described by periods in which it is} constant in time and equal to $\mathcal{D}^*$, {and periods in which it changes} according to an OU process with values $\mathcal{D}(t)>\mathcal{D^*}$. 
%Notably, other choices for the noise modulation do not significantly change our results.
%(see Methods).

We first consider the case in which the units are driven only by this extrinsic activity and not by the intrinsic one, i.e., we set the internal interactions to $\mathcal{W}_{ij} = 0$, $\forall i, j$. We refer to this case as the ``extrinsic model''. Then, we add back interactions by reconstructing an effective connectivity \cite{Gilson2016} in order to match the correlations matrix of our data, hence considering the ``interacting model''.

\subsection*{Scale-free avalanches from extrinsic activity}
\noindent In the absence of internal interactions, at each time-step the units are conditionally independent given the common external modulation $\mathcal{D}$. %Hence, we can write
%\begin{equation}
%    p(v_1, \dots, v_N, t \,|\, \mathcal{D}) = \prod_{i=1}^N p(v_i, t \,|\, \mathcal{D}).
%\end{equation}
However, we typically do not have {experimental} access to such external modulation. In general, we can only {describe the joint stationary probability distribution $p(v_1, \dots, v_N)$ of} the {units} alone. Let us now consider that $\gamma_{D} \gg \gamma_i$, which corresponds to the assumption that the time-scale of the modulation is much slower than the one of the units \cite{priesemann2018can}. In this limit, the process of $v_i$ reaches stationarity much faster than the process of $\mathcal{D}$, thus the joint stationary distribution is given by \cite{Nicoletti2021}
\begin{equation}
    p(v_1, \dots, v_N) = \int d\mathcal{D} \prod_{i=1}^N p(v_i\,|\, \mathcal{D}) p(\mathcal{D})
\end{equation}
{where $p(v_i|\mathcal{D})$ is the stationary solution to the Fokker-Planck equation\cite{gardiner2004handbook} associated to Eq.~\eqref{eqn:OUprocess_vi} at a fixed $\mathcal{D}$, and $p(\mathcal{D})$ is the stationary solution associated to Eq.~\eqref{eqn:OU_Dt}.}
Notice that, although the conditional probability distribution is factorizable, in general $p(v_1, \dots, v_N) \ne \prod_{i=1}^Np(v_i)$, i.e., the presence of the unobserved modulation results in an effective dependence between the units.

With the choice of an Ornstein-Uhlenbeck process for $v_i$ described in the previous section, $p(v_i | \mathcal{D})$ is Gaussian and we are able to compute these distributions analytically. We find
\begin{align}
\label{eqn:pstat_D}
    p(\mathcal{D}) = \frac{1}{2}\left[1+ \text{Erf}\left(\frac{\mathcal{D}^*}{\sqrt{\theta \gamma_D}}\right)\right]\delta(\mathcal{D}-\mathcal{D}^*) + \frac{H(\mathcal{D}-\mathcal{D}^*)}{\sqrt{\pi \theta \gamma_D}}e^{-\frac{\mathcal{D}^2}{\theta \gamma_D}},
\end{align}
where $H$ is the Heaviside step function, and
\begin{align}
\label{eqn:pstat_joint}
    p(v_i, v_j) = \frac{1+ \text{Erf}\left(\frac{\mathcal{D}^*}{\sqrt{\theta \gamma_D}}\right)}{2\pi \mathcal{D}^* \sqrt{\gamma_i\gamma_j}}e^{-\frac{1}{\mathcal{D}^*}\left(\frac{v_i^2}{\gamma_i}+\frac{v_j^2}{\gamma_i}\right)} + \frac{1}{\sqrt{\gamma_i\gamma_j\gamma_D \pi^3\theta}} \int_{\mathcal{D}^*}^\infty \frac{dD}{D} e^{-\frac{1}{D}\left(\frac{v_i^2}{\gamma_i}+\frac{v_j^2}{\gamma_i}\right)}e^{-\frac{D^2}{\theta \gamma_D}}.
\end{align}
As noted before, it is clear that $p(v_i, v_j) \ne p(v_i)p(v_j)$. In principle, we are able to compute the joint probability distribution for any number of variables in the same way. Crucially, notice that
\begin{equation}
\label{eqn:uncorrelated}
    \ev{v_i v_j} - \ev{v_i}\ev{v_j} = 0 \quad \forall i \ne j,
\end{equation}
which implies that the units, although not independent, are always uncorrelated. This follows immediately from the fact that all the expectation values where a variable $v_i$ appears an odd number of times vanish, e.g.,
\begin{align*}
 \ev{v_i v_j} = \int dv_i \,dv_j\, p(v_i, v_j) v_i\, v_j  =  \int d\mathcal{D} \, p(\mathcal{D}) \left(\int dv_i \, v_i\,p(v_i|\mathcal{D})\right)\left(\int dv_j\, v_j \,p(v_j|\mathcal{D})\right) = 0 \quad \forall i \ne j
\end{align*}
since $\int dv_i \,v_i \,p(v_i | \mathcal{D}) = 0$. Therefore, in the extrinsic {Ornstein-Uhlenbeck} model, the variables are always uncorrelated. This property will be useful when we will consider the case of a non-zero $\mathcal{W}_{ij}$, which will be the sole source of correlations in the model.

\begin{figure*}[t]
    \centering
    \includegraphics[width=\textwidth]{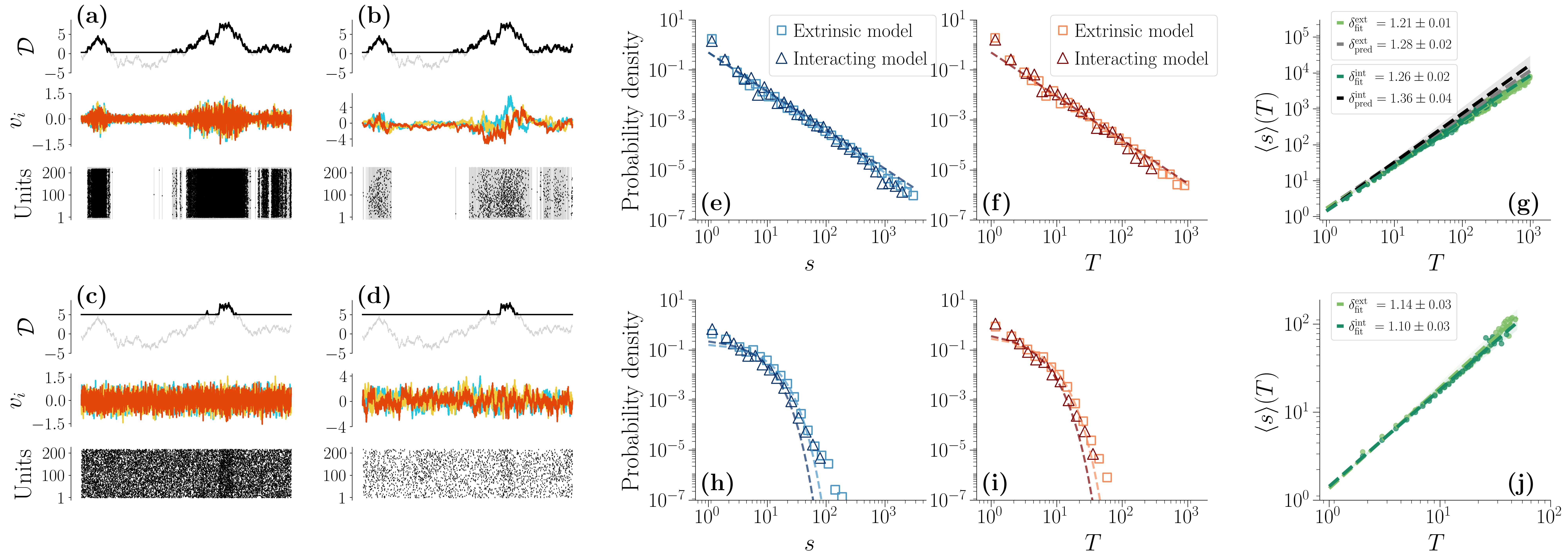}
    \caption{Avalanche statistics generated by the model at $\mathcal{D}^* = 0.3$ (a-b, e-g) and at $\mathcal{D}^* = 5$ (c-d,h-j), with $\gamma_D = 15$ and $\theta = 1$ and $\gamma_i = \gamma = 0.05$ for the extrinsic model. (a-b) Comparison between the trajectories of $\mathcal{D}(t)$, $v_i$ and the corresponding discretization in the low-$\mathcal{D}^*$ regime for (a) the extrinsic model and (b) the interacting one. (c-d) Same, but in the high-$\mathcal{D}^*$ regime. (e-g) If $\mathcal{D}^*$ is low, avalanches are power-law distributed with almost identical exponents in the extrinsic and interacting model, $\tau^\mathrm{ext} = 1.60 \pm 0.01$, $\tau^\mathrm{int} = 1.55 \pm 0.01$ and $\tau^\mathrm{ext}_t = 1.77 \pm 0.01$, $\tau^\mathrm{int}_t = 1.74 \pm 0.01$. The crackling-noise relation is verified in both cases. (h-j) Same plots, now in the high-$\mathcal{D}^*$ regime. Avalanches are now fitted with an exponential distribution. Notice that larger events, corresponding to periods in which $\mathcal{D}(t)>\mathcal{D}^*$, show up in the distributions' tails, suggesting that the shift between exponentials and power-laws is smooth. (j) The average avalanche size as a function of the duration scales with an exponent that, as $D^*$ increases, becomes closer to the trivial one $\delta_\mathrm{fit}^\mathrm{ext} \approx \delta_\mathrm{fit}^\mathrm{int} \approx 1$.}
    \label{fig:avalanches_model}
\end{figure*}

The dependence between the units induced by the modulation is shaped by the parameter $\mathcal{D}^*$. If $\mathcal{D}^*$ is high enough, the modulation is rare and the units are always dominated by noise as we can see in Figure \ref{fig:avalanches_model}c. On the other hand, if $\mathcal{D}^*$ is small, whenever $\mathcal{D}(t) = \mathcal{D}^*$ the noise contribution to the units will be vanishing and the activity will follow an exponential decay. Therefore, in this regime, each $v_i$ will typically alternate periods of quasi-silence to periods of activity. In other words, depending on the value of $\mathcal{D}^*$, this model can either reproduce a noise-driven behavior or a bursty, coordinated one, as shown in Figure \ref{fig:avalanches_model}a.

Most importantly, the low $\mathcal{D}^*$ regime is also the onset of power-law distributed avalanches of neural activity, as we see by simulating the model at different $\mathcal{D}^*$ and performing the same analysis as in LFPs. We find that, as $\mathcal{D}^*$ decreases, a transition between exponential decaying avalanches and power-law distributed ones appears. Figures \ref{fig:avalanches_model}e-f-g shows that for $\mathcal{D}^*$ small enough, the stochastic modulation produces scale-free avalanches in both size and time with exponents $\tau^\mathrm{ext} = 1.60 \pm 0.01$ and $\tau^\mathrm{ext}_t = 1.77 \pm 0.01$. Crucially, these avalanches satisfy the crackling-noise relation, that is we find $\delta_\mathrm{fit}^\mathrm{ext} = 1.21 \pm 0.01$ from the fit, while expecting $\delta_\mathrm{pred}^\mathrm{ext} = 1.28 \pm 0.02$ from the avalanches exponents. Moreover, the rescaled temporal profiles of these avalanches collapse to a single curve, {and the avalanche exponents is unaffected by the size of the system} (see Supplementary Information). These results are strictly related to a low value of $\mathcal{D}^*$, that is, on the alternating periods of low and high noise strength. In fact, for higher $\mathcal{D}^*$, the noise strength is always high and only exponential-decaying avalanches are present as we see in Figures \ref{fig:avalanches_model}h-i-j. {Notably, in this high-$\mathcal{D}^*$ regime, rare but large events that correspond to periods in which $\mathcal{D}>\mathcal{D}^*$ result in non-exponential tails of the distribution. This suggests that the shift between exponential and power-law avalanches is smooth, and indeed the avalanche exponents change gradually as $\mathcal{D}^*$ becomes smaller (see Supplementary Information).}

Let us note that these exponents are different from the ones obtained in LFP data, but this is perhaps not surprising. In fact, beside the simplicity of this paradigmatic model, such exponents have been found to depend on the experimental settings \cite{Fosque2021} and on individual variability \cite{Fontenele2019}. Nevertheless, our framework reproduces the scaling exponent $\delta \approx 1.28$ \cite{Fontenele2019, Buendia2021}, which we also find in our experimental settings in MUAs data \cite{Mariani2021b}. {Notably, in critical systems $\delta$ is expected to obey the} crackling-noise relation $\delta = (\tau_t -1)/(\tau -1)$ \cite{Sethna, Fontenele2019}. {However, following arguments similar to the ones proposed in \cite{Scarpetta}}, one can derive such relation {with the sole assumptions that avalanches are} power-law distributed and that they satisfy $s \sim T^\delta$, i.e., fluctuations in the size of an avalanche given its duration are negligible. Then,
\begin{equation}
    p(s(T)) \left|\frac{ds}{dT}\right|dT = p(T)dT \implies (T^\delta)^{-\tau} \delta T^{\delta-1} = T^{-\tau_t}
\end{equation}
from which it follows immediately that $\delta = (\tau_t -1)/(\tau -1)$. These assumptions are certainly satisfied in critical points, where the exponent $\delta$ is related to other critical exponents by a number of scaling relations \cite{Sethna}. Yet, the crackling-noise relation may hold also hold in other settings, as we find in our modeling framework.

\subsection*{Scale-free correlations from internal couplings}
We have shown that in the paradigmatic model described by Eq.~\eqref{eqn:OUprocess_vi} with $\mathcal{W}_{ij} = 0$ the extrinsic modulation alone generates power-law avalanches in the low-$\mathcal{D}^*$ limit. Yet, this extrinsic activity cannot explain correlations such as the ones observed in Figure \ref{fig:LFPData}e, {as shown by Eq.~\eqref{eqn:uncorrelated}}. Hence, we now consider the interacting model, i.e., we consider the case in which both the extrinsic and the intrinsic components of activity are present.

Since in the extrinsic model described in the previous section the units are uncorrelated, we can infer the values of $A_{ij}$ directly from the data \cite{Gilson2016}. In particular, we solve the inverse problem in such a way that the correlations of Eq.~\eqref{eqn:OUprocess_vi} match the experimentally-measured correlations $\sigma_{ij}$ of our LFPs (see Methods). The effective connectivity $A$ obeys the Lyapunov equation
\begin{equation}
\label{lyapunov}
\sum_{k}\left[\sigma_{ik} A_{kj} +A_{ik} \sigma_{kj} \right] = \delta_{ij}\int_{\mathcal{D}^*}^\infty \mathcal{D} \,p(\mathcal{D})\, d\mathcal{D} := \delta_{ij} f (\mathcal{D}^*, \gamma_D, \theta).
\end{equation} 
The different regimes for the interacting model with such interaction matrix are plotted in Figures \ref{fig:avalanches_model}b and \ref{fig:avalanches_model}d.

\begin{figure}[t]
    \centering
    \includegraphics[width=\textwidth]{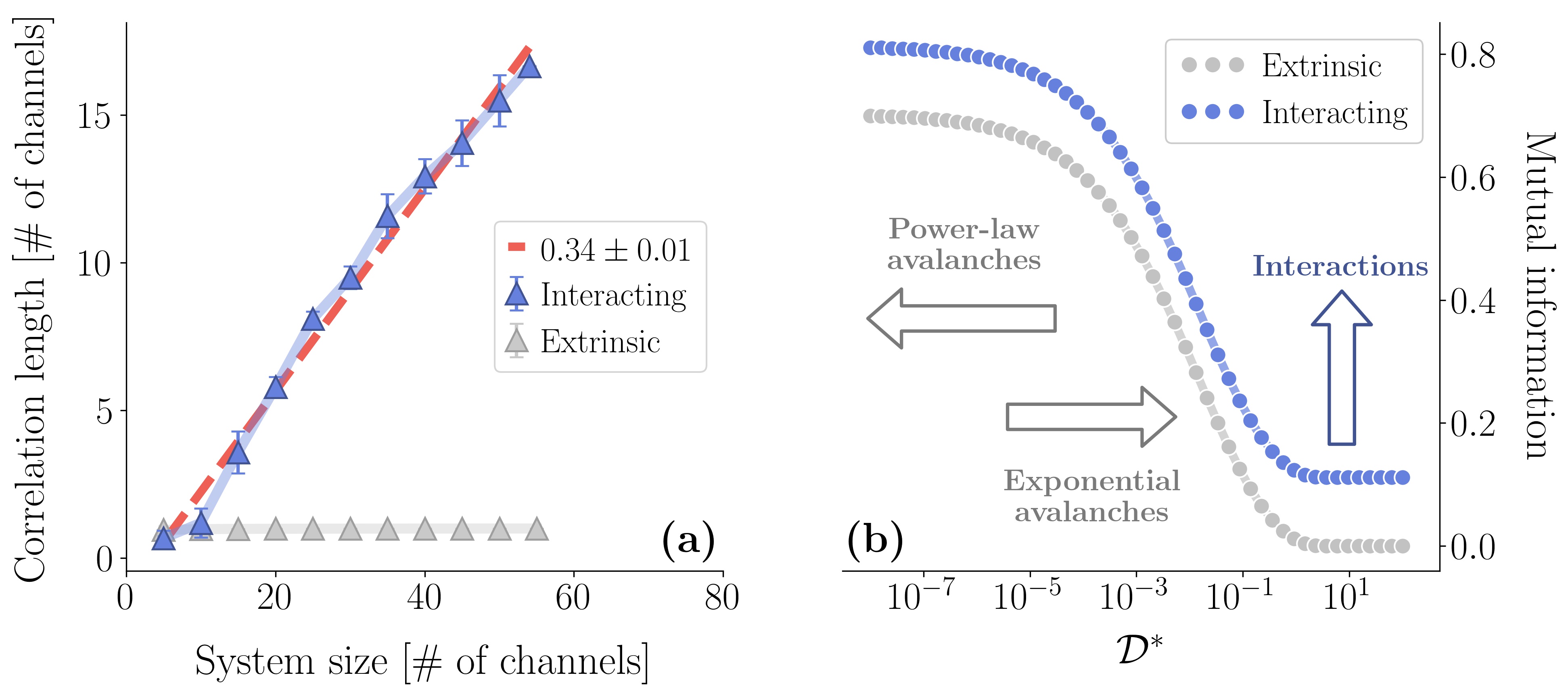}
    \caption{(a) The correlation length of the interacting model scales linearly with the system size, as in the data. In the extrinsic model, as expected, the correlation length of the fluctuations is constant and equal to 1, i.e., the correlation function drops to zero for adjacent electrodes. (b) Comparison between the mutual information in the extrinsic model ($\theta = 1$, $\gamma_D = 10$, $\gamma_1 = 0.1$, $\gamma_2 = 0.5$) and, as an example, in the interacting model with two units. Notice that the onset of a non-vanishing mutual information induced by $\mathcal{D}(t)$ is also the onset of power-law distributed avalanches, whereas the mutual information arising from interactions is independent of $\mathcal{D}^*$.}
    \label{fig:disentangling_correlations}
\end{figure}

As shown in Figures \ref{fig:avalanches_model}e-f-g, all avalanches exponents $\tau \approx 1.6$, $\tau_t \approx 1.75$ and the crackling-noise relation exponent $\delta \approx 1.28$ are not changed significantly by the inclusion of direct interactions among the units, nor the high $\mathcal{D}^*$ regime is changed either, as shown in Figures \ref{fig:avalanches_model}h-i-j. The fact that the exponents do not change when we add interactions to our model suggests that the avalanches are not affected by the interactions themselves - that is, in our model, they are determined by the extrinsic modulation. On the other hand, with these effective interactions we are now able to study the scaling of the correlation length $\xi$ as a function of the system sizes $L$. We plot the results in Figure \ref{fig:disentangling_correlations}a. As in our data, $\xi$ scales linearly with $L$ in the interacting model - in fact, we can show that, even though $A_{ij}$ depends on the modulation parameters $(\mathcal{D}^*, \gamma_D, \theta)$, the scaling of the correlation length does not (see Methods). Hence, in our model, scale-free correlations are not only strictly dependent on the interaction network, but they are completely unaffected by avalanches.

These results suggest a profound implication - in our model, extrinsic and intrinsic activity are disentangled. If we measure neural avalanches alone, we might not be able to infer anything about the intrinsic neural dynamics of our model. However, we might have access to such dynamics if we investigate the correlations of the fluctuations. This disentangling is deeply related to the dependency structure of our model. We can probe such structure by computing the mutual information
\begin{equation}
    \label{eqn:mutual}
    I = \int_{-\infty}^{+\infty} dv_i \int_{-\infty}^{+\infty} dv_j p(v_i, v_j) \log\frac{p(v_i,v_j)}{p(v_i)p(v_j)},
\end{equation}
which captures pairwise dependencies in the system that go well beyond simple correlations. In Figure \ref{fig:disentangling_correlations}b we show that a non-zero mutual information emerges in the extrinsic model in the low $\mathcal{D}^*$ limit. Remarkably, the onset of this dependency is also the onset of the coordinated behavior between the units, i.e., of power-law distributed avalanches. On the other hand, the mutual information vanishes only in the trivial limit $\mathcal{D}^* \to \infty$, since at finite $\mathcal{D}^*$ Eq.~\eqref{eqn:pstat_joint} is never exactly factorizable. When interactions are added back, the mutual information is simply shifted independently of $\mathcal{D}^*$ (see Methods), as we see in Figure \ref{fig:disentangling_correlations}b. Crucially, this interaction-dependent constant shift of the mutual information is a signal that the effects of external modulation and the effective interactions are completely disentangled, as recently shown for similar models \cite{Nicoletti2021}.

\subsection*{Scale-free avalanches in more biologically relevant extrinsic models}
The multivariate Ornstein-Uhlenbeck (mOU) is a paradigmatic model which, albeit simple enough to allow for an analytical treatment, does not account for many biological aspects. In order to show that our results generalize to more biologically sound models, we now consider the neural activity described by a Wilson-Cowan model \cite{Wilson, Benayoun, diSanto2018, wallace, decandia2021} as a variant of the extrinsic model. It includes both excitatory and inhibitory synapses and non-linearities in the transfer function, and its derivation is based on arguments over the dynamics of the neurons and action potentials \cite{Wilson}, which makes it a general tool to model mesoscopic neural regions.

We consider a stochastic version of the Wilson-Cowan model \cite{diSanto2018, Benayoun, decandia2021, wallace}, which includes a stochastic term that accounts for the finite size of the populations. We consider $N$ non-interacting neural populations, and each one is {modeled through the activity of two sub-populations, one of excitatory neurons $E_i$ and one of inhibitory neurons $I_i$. $E_i$ and $I_i$ are defined as the densities of active excitatory or inhibitory neurons, and can be interpreted as firing rates. They} evolve according to
\begin{equation}
\label{eq:wilson}
    \begin{cases}
        \frac{dE_i}{dt} = -\alpha E_i + (1-E_i) f(\omega_{E} E_i - \omega_{I} I_i + h) + \sqrt{(\alpha E_i + (1-E_i) f(\omega_{E} E_i - \omega_{I} I_i + h))}\eta_{E_i}\\
        \frac{dI_i}{dt} = -\alpha I_i + (1-I_i) f(\omega_{E} E_i - \omega_{I} I_i + h) + \sqrt{(\alpha I_i + (1-I_i) f(\omega_{E} E_i - \omega_{I} I_i + h))}\eta_{I_i}
    \end{cases}
\end{equation}
where $\alpha$ is the rate of spontaneous activity decay, $\omega_{E, I}$ are the synaptic efficacies, and $\eta_{E,I}$ are uncorrelated Gaussian white noises with {population-size dependent strength $\sigma \propto \frac{1}{\sqrt{K}} = \frac{I}{\sqrt{L}}$,with $K$ and $L$ that are the number of excitatory and inhibitory neurons corresponding to each neural population \cite{Benayoun, wallace}}. The response function $f(s)$ is given by
\begin{equation}
    \begin{cases}
        f(s) = {\beta} \tanh(s) & s \geq 0\\
        f(s) = 0 & s < 0
    \end{cases}
\end{equation}
where $s = \omega_{E} E_i - \omega_{I} I_i + h$ is the average incoming current from the other synaptic inputs and an external input $h$. For each unit, we are interested in the firing rate of the overall population $\Sigma_i = (E_i + I_i)/2$. {$\beta$ will be set to 1 from now on.} Importantly, we consider the case in which the units are inhibition dominated, i.e., when $\omega_{I} > \omega_{E}$, with a small noise amplitude $\sigma$, and are non-interacting with each other.

The external modulation comes into this model through the external current $h$. As a potential candidate for a biological realization of this external, stochastic driving, we consider the effective input that comes from other, yet unobserved, neural populations. Hence, we model $h$ as the firing rate $h = \left(E^{(h)} + I^{(h)}\right)/2$ of another Wilson-Cowan model, namely
\begin{equation}
\label{eq:wilsoninput}
    \frac{dh}{dt} = \frac{d}{dt}\left[\frac{E^{(h)} + I^{(h)}}{2}\right].
\end{equation}
{In order to reproduce avalanches, as the external input $h$ we should choose a stochastic modulation that displays bursts of activity separated by periods of silences. Importantly, it was recently shown in \cite{decandia2021} that this model admits a critical point at $\omega_{0_C} = \omega_E - \omega_I = \frac{\alpha}{\beta}$, where power-law distributed avalanches will emerge independently of the size of the system. Thus, clearly, a possible choice for the external stochastic modulation would be a Wilson-Cowan unit in the critical state.}

{Another potential candidate is still a Wilson-Cowan model, but in a balanced state \cite{Benayoun, diSanto2018} defined by $\omega^{(h)}_0 = \omega^{(h)}_E - \omega^{(h)}_I\ll \omega^{(h)}_{\rm S} = \omega^{(h)}_E + \omega^{(h)}_I$, and  we set the parameters so that $\omega^{(h)}_0 > \omega^{(h)}_{0_C}$. Crucially, in this scenario, the mechanism giving rise to avalanches is fundamentally different. With these parameters, and in the absence of noise, the dynamics predicts a stable up state. Yet, by increasing the noise amplitude, such up state can be destabilized, leading to large excursions in the down state and thus to avalanches. This phenomenon is a consequence of the non-normality of the matrix describing the linearized dynamics, that can cause a system to be reactive - i.e., its dynamics can exhibit unusually long-lasting transient behaviors even if it asymptotically converges to a stable fixed point, and that coincides with the condition $\omega_0 \ll \omega_S$ \cite{diSanto2018}.}

{For these reasons, here we choose as an effective input $h$ the firing rate coming from a neural population in a balanced state. In the Supplementary Information, we study the case in which such input comes instead from a population in the critical state $\omega_{0_C} = \frac{\alpha}{\beta}$, and we show that the results are qualitatively similar to the ones reported here.}

Intuitively, we are considering a case in which $N$ populations of neurons evolve according to a WC model in the inhibition dominated phase, and all receive the same input by another{, yet unobserved,} population of neurons.

\begin{figure*}[t]
    \centering
    \includegraphics[width=\textwidth]{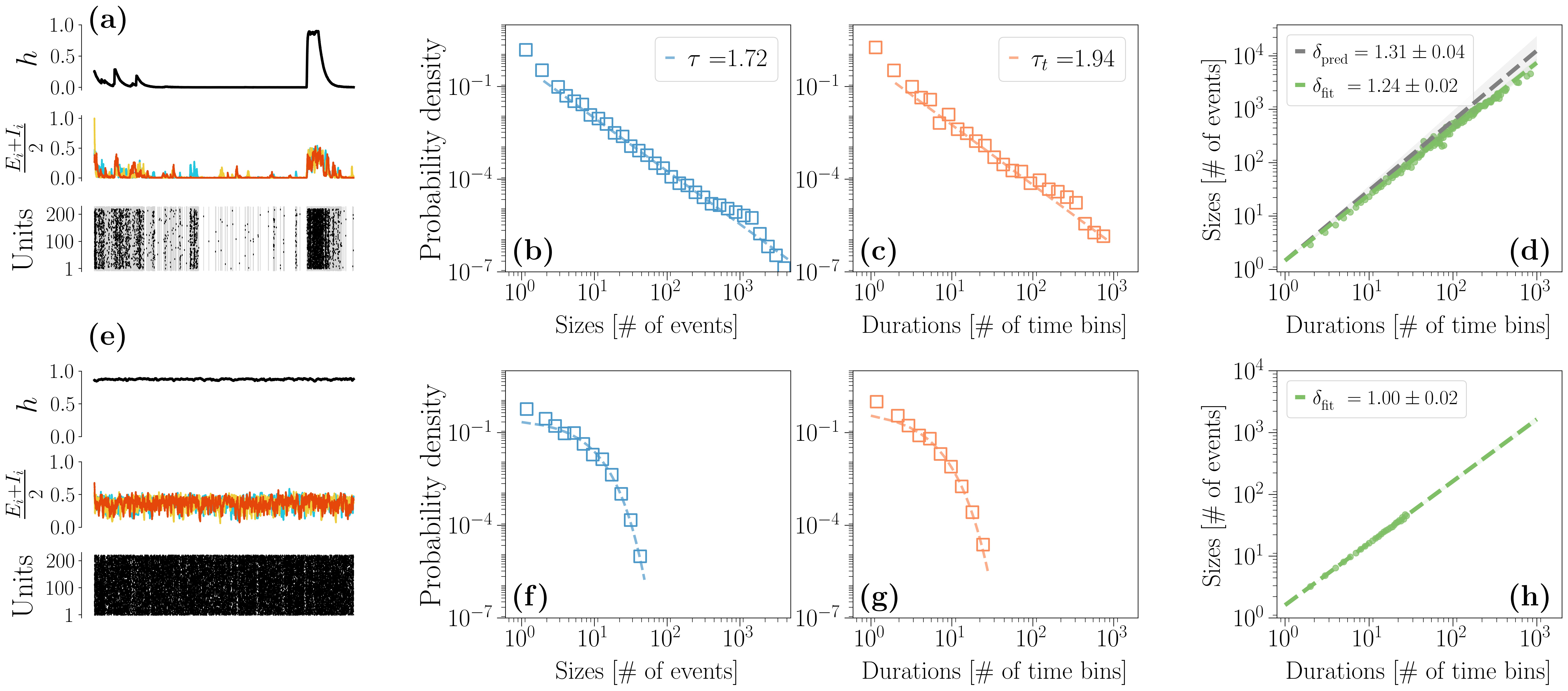}
    \caption{Avalanche statistics generated by the Wilson Cowan units. The Wilson Cowan units are always in an inhibition dominated phase, i. e. $\omega_{I} = 7$ and $\omega_{E} = 6.8$, and $\alpha = 1$. {Their external input $h$ is instead always in a balanced state, in particular $\omega^{(h)}_{E} = 50.5$, $\omega^{(h)}_{I} = 49.5$. Its other parameters are $h^{(h)} = 10^{-3}$ and $\alpha^{(h)} = 0.1$. In Figures (a-d) however, $\sigma^{(h)}$, the amplitude of the noise, is increased to $2.5 \times 10^{-2}$ so that the up state can be destabilized by the noise. In Figures (e-h) instead the noise is reduced to $5 \times 10^{-3}$ so that the up state is stable.}
    (a, e) Comparison between the trajectories of $h$, $\frac{E_i + I_i}{2}$ and the corresponding trains of events in the high (a) and low (e) $\sigma^{(h)}$ regime. (b-d) If $\sigma^{(h)}$ is high avalanches are power-law distributed and the crackling-noise relation is verified. (f-g) Same plots, now in the low $\sigma^{(h)}$ regime. Avalanches are now fitted with an exponential distribution. (h) The average avalanche size as a function of the duration scales with an exponent that, as $\sigma^{(h)}$ decreases, becomes closer to the trivial one $\delta_\mathrm{fit} \approx 1$.}
    \label{fig:avalanches_WC}
\end{figure*}

Although we cannot analytically tackle this model, we simulate the Langevin equations Eqs.~\eqref{eq:wilson}-\eqref{eq:wilsoninput} and from each firing rate $\Sigma_i$ we generate trains of events and analyze avalanches by temporal binning through the average inter-event interval (see Methods). If the noise strength $\sigma_h$ is high enough, $h$ spends most of the time close to the down state, while showing frequent bursts of activity - this behavior is qualitatively equivalent to the low-$\mathcal{D}^*$ limit previously considered, and is depicted in Figure \ref{fig:avalanches_WC}a. Again, in this regime we find that avalanche sizes and durations are power-law distributed and satisfy the crackling-noise relation, as we see in Figure \ref{fig:avalanches_WC}b-d. On the other hand, if $\sigma_h$ is lower, the external input $h$ is not modulated as in the previous high $\mathcal{D}^*$ limit. Once more, {in this regime} avalanches obtained through temporal binning are exponentially suppressed both in their size and in their duration, as we see in Figure \ref{fig:avalanches_WC}e-h.

Hence, the proposed extrinsic mechanism is valid beyond the paradigmatic case of an Ornstein-Uhlenbeck process we previously considered. Finally, we note that a balanced $h = \left(E^{(h)} + I^{(h)}\right)/2$ is only one of the possible choices that could be considered for the external modulation. {Crucially, we believe it is one that achieves a significant biological realism, being linked to E-I balance \cite{Poil2012}. Moreover, a large number of parameters can satisfy this condition, thus not requiring extreme fine-tuning, as long as the size of the system remains finite (e.g., for our choice of parameters, at a noise level $\sigma^{(h)} = 5 \times 10^{-3}$ corresponding to $K$ = $L =  40000$ neurons, avalanches are not present \cite{Benayoun}). We highlight that other choices of external modulation should be able to generate bursts of activity separated by periods of silence. For instance, another relevant choice would be Brunel's model in the synchronous irregular phase \cite{Brunel}.}

\section*{Discussion}
We have measured the activity in spatially-extended LFPs data from the rat's barrel cortex and we have found the appearance of neural avalanches and of the crackling noise relation. However, a variety of mechanisms can generate these properties.
Nonetheless, our data also display a correlation length that scales linearly with the size of the system, a key feature of critical systems. In order to understand the possible origin of these signatures of criticality, we have developed a archetypal, but of analytical ease, framework where the intrinsic contributions to the neuronal activity - due to the direct interaction between the units themselves - and the extrinsic ones - arising from externally-driven modulated activity - are exactly disentangled.
{Crucially, our work and our results fit in a well-established and fruitful research line that studies null mechanisms for the emergence of neuronal avalanches \cite{Touboul2017, priesemann2018can}. We believe that being able to disentangle such null mechanisms with more biological insightful properties (e.g. neural correlations) is instrumental in understanding what avalanches can teach us about neuronal and brain dynamics. Thus, building on recent works that have stressed the importance of considering external inputs when studying brain criticality \cite{Fosque2021}, our work wants to highlight the importance of disentangling extrinsic and intrinsic factors, that inherently contribute to neural activity.
}

{When considering a properly chosen external modulation}, our model displays a regime in which power-law avalanches that satisfy the crackling-noise relation emerge and that are compatible with the exponent $\delta \approx 1.28$ {found in}\cite{Fontenele2019, Buendia2021}. Crucially, this result holds even in other extrinsic models, such as when considering Wilson-Cowan units {modulated by an effective input coming from unobserved neural populations}. {The same value of the $\delta$ exponent is known to have been found} in a variety of neural systems, and our results suggest that it could be explained by a slow time varying extrinsic dynamics \cite{priesemann2018can} that affects all the neural units in the same way. {On the other hand, it was recently shown \cite{Carvalho2021} that this exponent may arise as a consequence of measuring only a fraction of the total neural activity, i.e., of subsampling. Let us also note that avalanches distributed with all exponents compatible with a critical branching process (i. e., with $\delta = 2$) were found experimentally in \cite{Miller2019}, once properly taking into account the role of gamma-oscillations. Further work is still needed to understand the emergence of such exponents and in which conditions they are robustly reproduced in experiments.}

At the same time, while scale-free spatial correlations can and do coexist with power-law avalanches, these kinds of critical signatures cannot be explained by the extrinsic activity alone. Crucially, our archetypal model allows us to combine this extrinsic dynamics to an intrinsic interaction matrix, inferred directly from the experimental data to match the spatial correlations we find in our experiments. When we do so, we show that these two signatures of criticality can be disentangled - {avalanches appear as a consequence of the external modulation and are only slightly affected by the interactions}, and, vice-versa, the interactions determine the spatial correlations independently of the external modulation. Hence, we believe that scale-free correlations may be deeply related to the origin of criticality in the brain, playing a fundamental role in the advantages it might achieve by being critical \cite{kinouchi2006optimal, Munoz2018, Bialek2011}.

Remarkably, it was recently shown that in models similar to the one considered here the mutual information always receives distinct and disentangled contributions from the internal interactions and from an environmental, external dynamics \cite{Nicoletti2021}. Although, clearly, the presence of a non-zero mutual information cannot be a sufficient condition for power-law avalanches to appear, in our extrinsic model their emergence does correspond to the onset of a non-vanishing mutual information. This fact suggests a promising future perspective. By explicitly considering both the intrinsic activity and the extrinsic contributions, one might be able to combine all these considerations into a unified information-theoretic view - perhaps helping to unfold the underlying biological mechanisms at the origin of the observed signatures of criticality in neural activity.

\section*{Methods}
{The study is reported in accordance with ARRIVE guidelines.}
\subsection*{Experimental setting}
\subsubsection*{Surgical procedures}
LFPs recordings are performed on Wistar rats, which are maintained under standard environmental conditions in the animal research facility of the Department of Biomedical Sciences of the University of Padova. All the procedures are approved by the local Animal Care Committee (O.P.B.A.) and the Italian Ministry of Health (authorization number 522/2018-PR) and all methods are performed in accordance with relevant guidelines and regulations. Young adult rats aged 36 to 43 days and weighting between 150 and 200 g are anesthetized with an intra-peritoneal induction mixture of tiletamine-xylazine (2 mg and 1.4 g/100 g body weight, respectively), followed by additional doses (0.5 mg and 0.5 g/100 g body weight) every hour. The anesthesia level is constantly monitored by testing the absence of eye and hind-limb reflexes and whiskers’ spontaneous movements. Each animal is positioned on a stereotaxic apparatus where the head is fixed by teeth- and ear-bars. To expose the cortical area of interest, an anterior-posterior opening in the skin is made in the center of the head and a window in the skull is drilled over the somatosensory barrel cortex at stereotaxic coordinates $-1 \divisionsymbol -4$ AP, $+4 \divisionsymbol +8$ ML referred to bregma \cite{swanson2003}.  A slit in the meninges is then carefully made with fine forceps at coordinates $- 2.5$ AP, $+ 6$ ML for the subsequent insertion of the recording probe. As a reference, the depth is set at 0 µm when the electrode proximal to the chip tip touches the cortical surface. The neuronal activity is recorded from the entire barrel cortex (from 0 to $- 1750$ $\mu m$), which is constantly bathed in Krebs’ solution (in mM: NaCl 120, KCl 1.99, NaHCO\textsubscript{3} 25.56, KH\textsubscript{2}PO\textsubscript{4} 136.09, CaCl\textsubscript{2} 2, MgSO\textsubscript{4} 1.2, glucose 11). An Ag/AgCl electrode bathed in the extracellular solution in proximity of the probe is used as reference.

\subsubsection*{Recordings}
LFPs are recorded through a custom-made needle that integrates a high-density array, whose electrodes are organized in a $64\times4$ matrix. The operation principle of the multi-electrode-arrays used to record LFPs is an extended CMOS based EOSFET (Electrolyte Oxide Semiconductor
Field Effect Transistor). The recording electrodes are 7.4$\mu$m in diameter size and the needle is 300 $\mu$m in width and 10$mm$ long. The x- and y-pitch (i.e. the distance between adjacent recording sites) are 32$\mu$m. The multiplexed signals are then digitized by a NI PXIe-6358 (National Instruments) up to 1.25MS/s at 16bit resolution and saved to disk by a custom LabVIEW acquisition software.
The LFP signal is sampled at 976.56 Hz and band-pass filtered (2-300 Hz). The dataset analyzed for this work consist in 20 trials of basal activity lasting 7.22 seconds, that are recorded from 4 rats.

\subsection*{LFPs peaks detection}
For the detection of LFP events, the standard deviation (SD) and the mean of the signal were computed for each channel. In order to distinguish real events from noise, a three SD threshold was chosen based on the distribution of the signal amplitudes which significantly deviated from a Gaussian best fit above that threshold. Both negative and positive LFPs (i.e., nLPFs and pLFPs, respectively) were considered as events in accordance with previous works \cite{Shew2015}. One reason is that across the depth of the cortex there are polarity changes in the LFP signal because of compensatory capacitive ionic currents, particularly along dendrites of pyramidal cells \cite{Buzsaki}. Since in our experiments electrodes span multiple cortical layers, both nLFPs and pLFPs were found and detected. Moreover, alternatively, pLFPs can be related to the activation of populations of inhibitory neurons. %leading to inhibitory outward postsynaptic currents, which also justifies their inclusion in the events count. 
For detection, each deflection was considered terminated only after it crossed the mean of the signal.

\subsection*{Avalanches analysis}
To study avalanches' statistics, the data are temporally binned, and avalanches are defined as sequences of bins that present activity, and an avalanche ends once an empty bin is found - the temporal bin chosen is the average inter-event interval \cite{beggs2003}. Then, the distribution $p(s)$ of the avalanches sizes - the number of events in each avalanche - and of the avalanche duration $p(T)$ are computed and fitted using a corrected maximum likelihood method \cite{Gerlach2019, Mariani2021b}. In particular, following the methods proposed in \cite{Clauset, Gerlach2019}, avalanche sizes and lifetimes are fitted with discrete power-laws $p(y; \alpha) = \frac{y^{-\alpha}}{\sum_{x = x_{min}}^{x = x_{max}}x^{-\alpha}}$. The parameter $x_{max}$ is set to the maximum observed size or duration. $x_{min}$ is selected as the one that minimizes the Kolmogorov-Smirnov distance (KS) between the cumulative distribution function (CDF) of the data and the CDF of the theoretical distribution fitted with the parameter that best fits the data for $y \geq x_{min}$ \cite{Clauset}.
To assess goodness-of-fit we compared the experimental data against 1000 surrogate datasets drawn from the best-fit power-law distribution with the same number of samples as the experimental dataset. The data were considered power-law distributed if the fraction of the KS statistics of the surrogates which were greater than the KS statistic for the experimental data was greater than 0.1.
We also take into account the fact that while maximum likelihood methods rely on the independence assumption, actual data often display correlations, and this may lead to false rejection of the statistical laws \cite{Gerlach2019}. As the authors of \cite{Gerlach2019} suggest, before performing the fit and assessing p-values, we undersample the data in order to decorrelate them, by  estimating the time $\tau^*$ after which two observations (e.g., the avalanche sizes) are independent from each other, as done in \cite{Gerlach2019, Mariani2021b}.

\subsection*{Scaling of the correlation length}
\noindent The correlation length $\xi$ can be defined as the average distance at which the correlation of the fluctuations around the mean crosses zero \cite{Cavagna2010}, and it is known to diverge at criticality in the thermodynamic limit \cite{binney1992}. For finite systems, this behavior can be probed by computing the correlation length at different system sizes. For each time-series we compute their fluctuations around the mean activity,
\begin{equation}
    \Tilde{v}_i(t) = v_i(t) - \frac{\sum_{i= 1}^Nv_i(t)}{N}.
\end{equation}
Different sizes of the system, i.e., different portions of the array, are selected and, importantly, the mean activity is computed for each system size, considering the channels inside the portion of the array \cite{martin2020}. Hence, we study the behavior of the correlation length with system sizes corresponding to different subsamples from the multi-electrodes array probe \cite{Ribeiro2020}. We assume that the units of our model have the same topology as our data, i.e. we assume that the units are placed as the channels in the $55 \times 4$ array of our experimental setup. Since in our case the array shape is rectangular, we consider the number of rows $L$ as the relevant dimension and build subsampled systems of size $L \times 4$, with $L$ that decreases from the maximum of $55$ channels down to $5$ channels.

For each system's subset, we compute the average correlation function of the fluctuations between all pairs of channels separated by a distance $r$,
\begin{equation}
    {C(r)} = \ev{\frac{\ev{\left(\tilde{v}_{i}-\overline{\tilde{v}}_{i}\right)\left(\tilde{v}_{j}-\overline{\tilde{v}}_{j}\right)}_{t}}{\sigma_{\tilde{v}_{i}} \sigma_{\tilde{v}_{j}}}}_{i, j}
\end{equation}
where $\ev{\cdot}_t$ stands for the average over time, $\ev{\cdot}_{i,j}$ is the average over all pairs of channels separated by a distance $r$ and
\begin{gather*}
    \overline{\tilde{v}}_{i} =\frac{1}{T} \sum_{t=1}^{T} \tilde{v_i}\left(t\right) \\
\sigma_{\tilde{v}_{i}}^{2} =\frac{1}{T} \sum_{t=1}^{T}\left(\tilde{v_i}\left(t\right)-\overline{\tilde{v}}_{i}\right)^{2}
\end{gather*}
with $T$ is the length of the time series. Then $\xi$ is computed as the zero of the correlation function $C(r = \xi) = 0$. To reduce the noise effects, results were averaged across all possible sub-regions for any given size. Then the $\xi$ are plotted against the relative system size $L$ and the slope of the fit is obtained through linear regression.

\subsection*{Correlations in the interacting model}
\noindent The process studied in the main text is a multivariate Ornstein-Uhlebeck process \cite{gardiner2004handbook} of the form
\begin{equation}
    d \boldsymbol{v}(t)=-A \boldsymbol{v}(t) d t+ B(t)d \boldsymbol{Z}(t),
    \label{Lyapu_EQ}
\end{equation}
where $B(t)$ is a diagonal matrix whose diagonal elements are given by $\sqrt{\mathcal{D}(t)}$ {and $\boldsymbol{Z}(t)$ denotes a Wiener process}. In the case of non-interacting units, which we use to model the extrinsic activity, the matrix $A$ is again diagonal with entries $A_{ij} = \delta_{ij}/\gamma_i$. If the matrix $B$ were constant in time, the covariance matrix $\sigma$ of the mOU would be determined by the continuous Lyapunov equation \cite{gardiner2004handbook, Gilson2016}
\begin{align}
    A \sigma +\sigma A^{\mathrm{T}}= B B^{\mathrm{T}}.
\end{align} 
Since in our case the matrix $B$ is a stochastic variable, we need to marginalize over its stationary distribution $p(B)$. Then, we immediately get
\begin{equation}
A \sigma +\sigma A^{\mathrm{T}}=  Q, 
\label{l_eq}
\end{equation} 
where $Q$ is a diagonal matrix whose elements are given by
\begin{equation}
    Q_{ij} = \delta_{ij}\int_{\mathcal{D}^*}^\infty \mathcal{D} \,p(\mathcal{D})\, d\mathcal{D} := \delta_{ij} f (\mathcal{D}^*, \gamma_D, \theta).
    \label{Q_eq}
\end{equation}
Then, taking the transpose of Eq. (\ref{l_eq}) and assuming that $A$ is symmetric, we end up with a Lyapunov equation for the matrix $A$ -  $\sigma A + A \sigma =  Q$. 
In principle, we could relax the assumption of symmetry of the matrix $A$ by considering the covariance matrix and the time shifted covariances \cite{Gilson2016}. However, this introduces further approximations and in the present work we are only interested in the covariance matrix. Hence we end up with a model
\begin{equation}
    \dot{v}_i(t) = -\sum_j A_{ij}v_j(t) + \sqrt{\mathcal{D}(t)} \xi_i(t)
\end{equation}
where $A_{ij}$ depends on the the parameters of the stochastic modulation $(\mathcal{D}^*, \gamma_D, \theta)$. Notice that if we write $\tilde{A}_{ij} = A_{ij}/f(\mathcal{D}^*, \gamma_D, \theta)$ we need to solve the Lyapunov equation $\sigma \tilde A + \tilde A\sigma = \mathbb{1}$ that only depends on $\sigma$, the covariance matrix of the data. Then, if we rescale the experimental time series by their standard deviation, $\sigma$ coincides with the correlation matrix of our data, and we refer to it in the main text. If we introduce $\tilde{\mathcal{D}} = \mathcal{D}/f$ and $\tilde{v}_i = v_i/\sqrt{f}$ we end up with
\begin{equation*}
    \dot{\tilde v}_i(t) = -\sum_j \tilde{A}_{ij} \tilde v_j(t) + \sqrt{\tilde{\mathcal{D}}(t)} \xi_i(t)
\end{equation*}
and clearly
\begin{equation*}
    \ev{\tilde v_i \tilde v_j} - \ev{\tilde v_i}\ev{\tilde v_j} = \frac{\ev{ v_i  v_j} - \ev{v_i}\ev{v_j}}{f}.
\end{equation*}
Therefore, the covariance between $\tilde v_i$ and $\tilde v_j$ is proportional to the covariance between $v_i$ and $v_j$. This implies that at different $(D^*, \theta, \gamma_D)$ we find a rescaled interaction matrix $A_{ij}$, but the scaling of correlation length does not change.

\subsection*{Mutual information in the interacting model}
The stationary probability distribution solution of the interacting model with a generic interaction matrix $A$ is
\begin{equation}
    p(v_1, \dots, v_N) = \frac{1 + \mathrm{Erf}\left[\frac{\mathcal{D}^*}{\sqrt{\theta\gamma_D}}\right]}{2\sqrt{(\pi\mathcal{D}^*)^N \det \Sigma}} e^{-\frac{1}{\mathcal{D}^*}\vb{v}^T {\Sigma}^{-1}\vb{v}} + \frac{1}{\sqrt{(\gamma_D\theta)^N \pi^{N+1}\det \Sigma}}G_N\left(\frac{\mathcal{D}^*}{\sqrt{\theta \gamma_D}}, \frac{\vb{v}^T {\Sigma}^{-1}\vb{v}}{\sqrt{\theta \gamma_D}}\right) 
\end{equation}
where the matrix $\Sigma$ is determined by $\left({A}{\Sigma} + {\Sigma}{A}^T \right)/2 = \mathbb{1}$ and
\begin{equation}
    G_N(\alpha, \beta) = \int_{\alpha}^{\infty} \frac{dx}{x^{N/2}}e^{-\frac{\beta}{x}-x^2}.
\end{equation}
In general, we can define a multivariate information between these $N$ variables and the results of the main text would not change. In practice, however, it is very hard to perform the related numerical integration if $N$ is large. Therefore, in the main text we consider the exemplary case of the mutual information, i.e., the case of two variables that interact through the matrix element $A_{12} = \tilde{A}_{12} f(\mathcal{D}^*, \gamma_D, \theta)$ with $\tilde{A}_{12} = 2$. As recently shown for simpler models \cite{Nicoletti2021}, the mutual information of this model receives a distinct - and constant - contribution from the interaction matrix and from the modulation induced by $\mathcal{D}^*$. In fact, if we consider $A = \mathrm{diag}(\gamma_1, \dots, \gamma_N)$, i.e., we consider the extrinsic model, the only the contribution to the mutual information comes from the external modulation, whereas the addition of interactions simply shifts the mutual information by a constant value at all $\mathcal{D}^*$.

\section*{Availability of materials and data}
The dataset generated for this study is available upon request to corresponding author.
%\bibliography{paper_bib}

\section*{Acknowledgements}
S.S. acknowledges DFA and UNIPD for SUWE\_BIRD2020\_01 grant, and INFN for LINCOLN grant. S.V. acknowledges support from Horizon 2020 (European Commission), FET Proactive, SYNCH project (GA number 824162).
We thank Claudia Cecchetto for the initial analyses on the electrophysiological data and suggestions on relevant literature and Victor Buendia for insightful discussions.

\section*{Author contributions statement}
S.S. designed the theoretical study and supervised the research with S.V., while M.B. and S.V. designed the experiments. M.M. conducted the experiments. B.M. analyzed the data. B.M., G.N. and S.S. formulated the model. B.M. and G.N. performed the analytical calculations, implemented the simulations and analyzed the results. B.M., G.N., S.V. and S.S. interpreted the results. G.N. prepared the figures. All authors contributed to the article and approved the submitted version. 

\section*{Competing interests}
The authors declare no competing interests.

\newpage
%\pagestyle{empty}
%%%%%%%%%% Merge with supplemental materials %%%%%%%%%%
\pagebreak
%\begin{widetext}

%%%%%%%%%% Merge with supplemental materials %%%%%%%%%%
%%%%%%%%%% Prefix a "S" to all equations, figures, tables and reset the counter %%%%%%%%%%
\setcounter{equation}{0}
\setcounter{figure}{0}
\setcounter{table}{0}
\setcounter{page}{1}
\setcounter{section}{0}
\setcounter{subsection}{0}
\makeatletter
\renewcommand{\theequation}{S\arabic{equation}}
\renewcommand{\thefigure}{S\arabic{figure}}
\renewcommand{\thesection}{S\Roman{section}}

\begin{center}\Large{Supplementary Information: Disentangling the critical signatures of neural activity}\end{center}

\maketitle
\section*{Additional avalanches statistics in the model}
\noindent Another signature of criticality is the collapse of the average profile of avalanches of widely varying duration onto a single scaling function. For avalanches of duration $T$ we can write down the average number of firing at time $t$ as $s(t,T) = T^{\delta-1} F(t/T)$ where F is a universal scaling function that determines the shape of the average temporal profile. $\ev{S(T)}$ and $s(t,T)$ are related by $\ev{S(T)} = \int_0^Ts(t,T)dt$. At the critical point we expect that plots of $t/T$ versus $s(t,T)T^{1-\delta}$ for different $T$ will collapse onto the same universal scaling function \cite{Sethna}.

Thus, finding the exponent for which the goodness of the collapse is higher provides another way to estimate $\delta$. For testing the avalanche shape collapse, we used the methodology introduced in \cite{Timme2016}. To determine the quality of the collapse, the averaged and rescaled avalanche profiles of different lifetimes $F(t/T) = T^{1-\delta}s(t/T,T) $ are first linearly interpolated at $1000$ points along the scaled duration. The variance across the different $F(t/T)$ is calculated at each interpolated point, and the shape collapse error $\epsilon(\delta)$ is then defined as the mean variance divided by the squared span of the avalanche shapes, where the span equals the maximum minus the minimum value of all rescaled avalanche profiles. In the presented analysis, avalanche shapes of  $T>10$ bins with at least 10 samples were used.

\begin{figure}[h!]
    \centering
    \includegraphics[width=12cm]{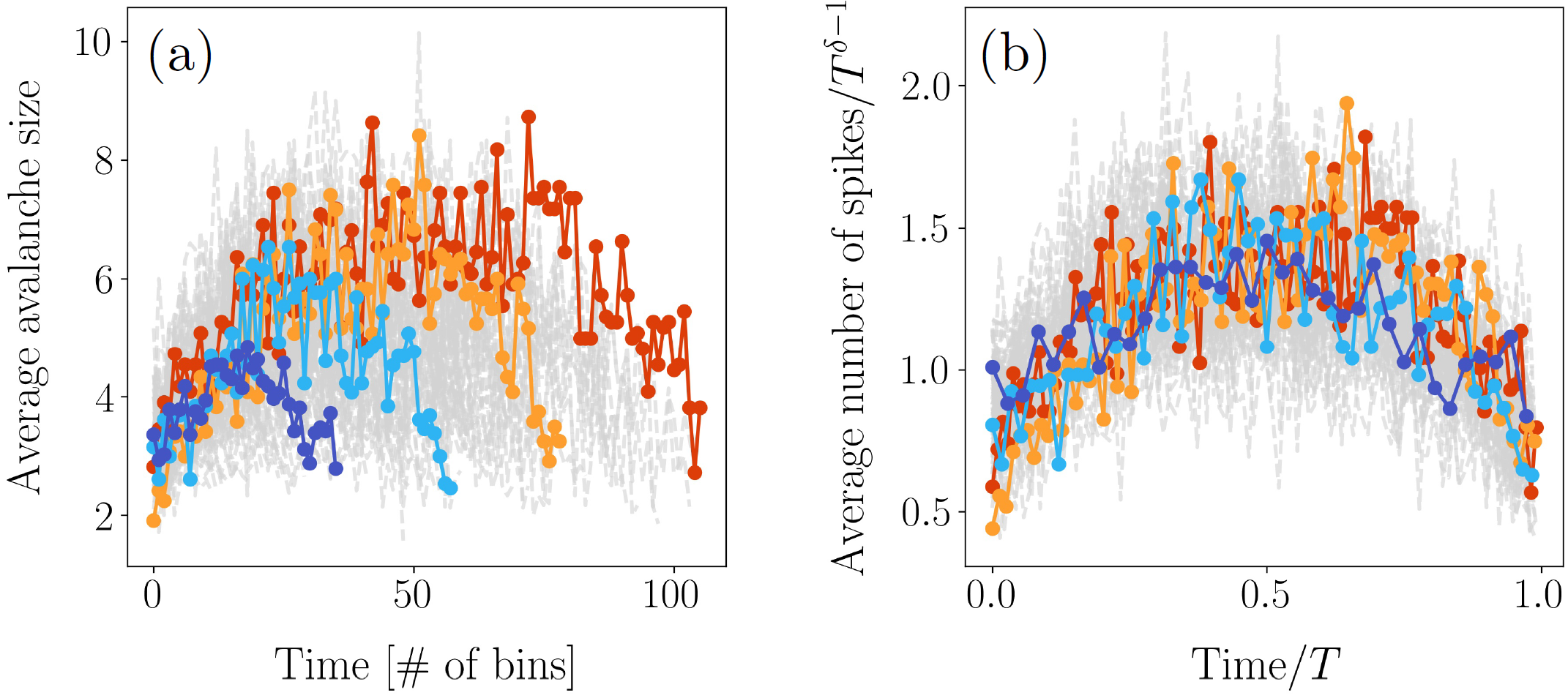}
    \caption{Collapse of the average profile of avalanches of varying duration in the extrinsic model, for the low $\mathcal{D}^*$ regime. (a) Profile of the avalanches before the rescaling. (b) If we rescale with an exponent $\delta \approx 1.33$, which is remarkably close to the one found in the main text through the crackling-noise relation, we obtain an optimal collapse onto the same scaling function.}
    \label{fig:collapse}
\end{figure}

The collapse has been tested on the extrinsic model in the case of low $\mathcal{D}^*$, and the results are plotted in Figure \ref{fig:collapse}. We find that the exponent that minimizes $\epsilon(\gamma)$ turns out to be $\approx 1.33$, close to the estimates of $\delta $ found through the linear fit of average size given duration and through the prediction of the crackling-noise relation in the main text. Again, it is also close to the apparently super-universal exponent found in \cite{Fontenele2019} and in \cite{Friedman2012}. 
% Finally, we found that the average profile collapse on the scaling function has the form of an inverted parabola, as it has been found also in other experiments \cite{Friedman2012}, and analogously in models of inhomogeneous Poisson processes \cite{Touboul2017}.

\section*{Testing the predictions of the extrinsic model}
\noindent Here, we show that simulations of the extrinsic model described by an Ornstein-Uhlenbeck process agree with the analytical results presented in the main text. Whenever not specified, we assume that the parameters of the model are given by $\mathcal{D^*} = 0.3$, $\theta = 1$, $\gamma_D = 10$ together with $\gamma_i = 0.1$, $\gamma_j = 0.5$. Thus, we are in the limit of timescale separation considered in the main text.

\begin{figure}[h!]
    \centering
    \includegraphics[width = 0.9\textwidth]{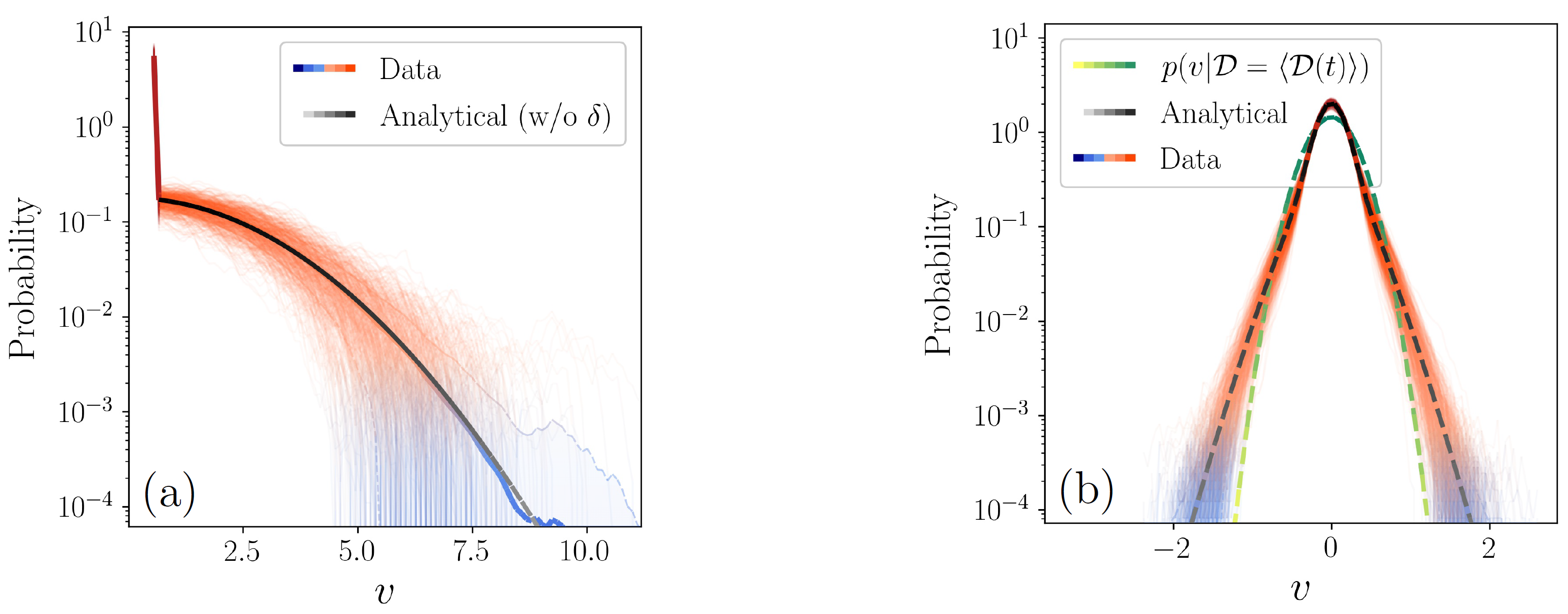}
    \caption{Comparison between the stationary distributions obtained in the main text and the results $10^3$ simulations. Semi-transparent lines represent different simulations. Filled areas of the plots represent one standard deviation from the mean distribution. (a) Probability distribution of $\mathcal{D}$. (b) Probability distribution of a single $v_i$.}
    \label{fig:probability_single}
\end{figure}

\begin{figure}[h!]
    \centering
    \includegraphics[width=12.6cm]{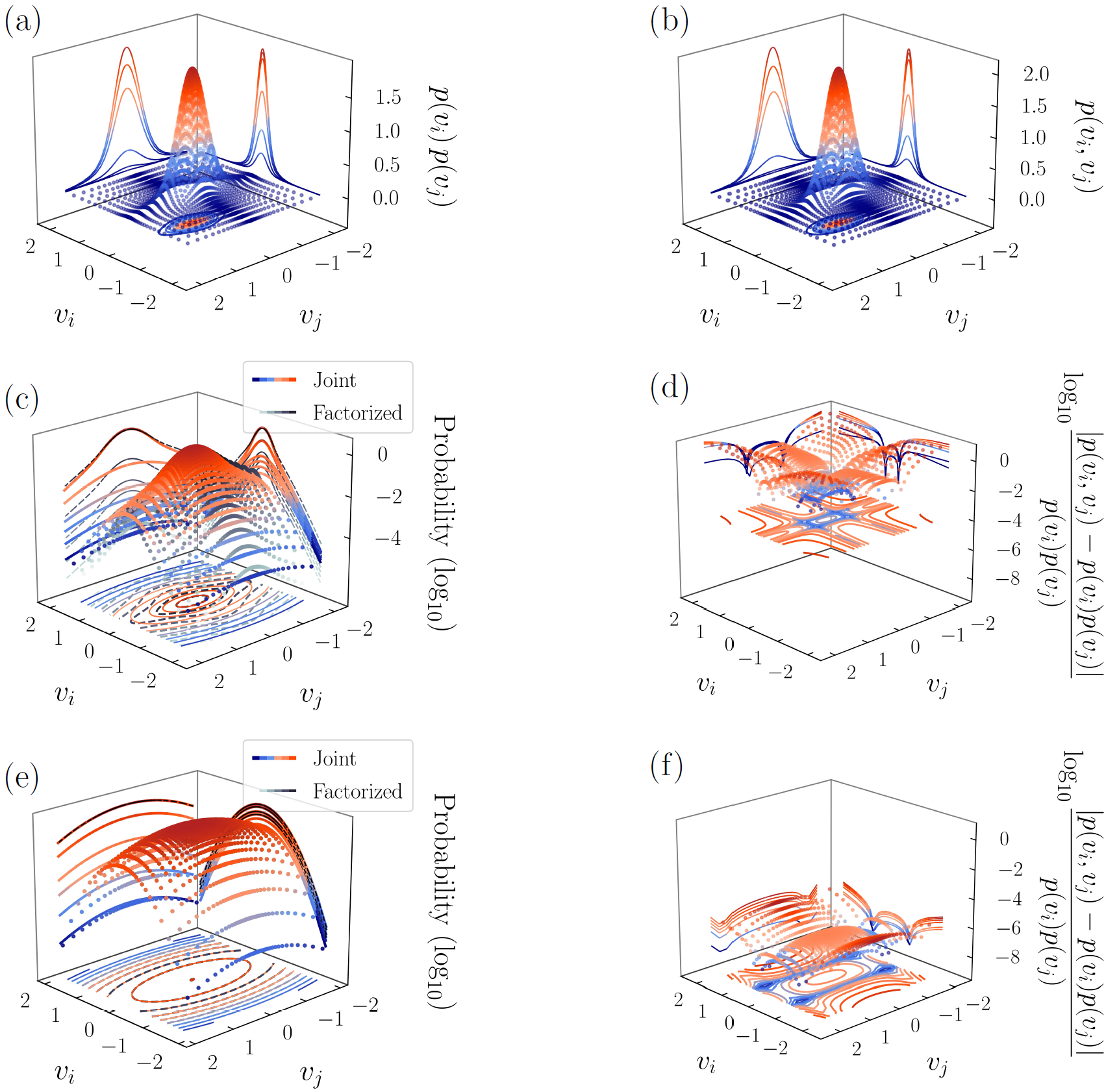}
    \caption{Comparison between the analytical expressions of the joint probability distribution in Eq. (\ref{eqn:pstat_joint_si}) and its factorization $p(v_i, v_j)$. (a) The factorized distribution. (b) The joint distribution. (c) Comparison between the two in a log-plot. (d) Relative difference between the two with respect to the factorized distribution. (e-f) Comparison between the joint and the factorized distributions, with the same parameters except for $\mathcal{D}^* = 5$.}
    \label{fig:probability_joint_factor}
\end{figure}

\begin{figure}[t!]
    \centering
        \includegraphics[width = 11cm]{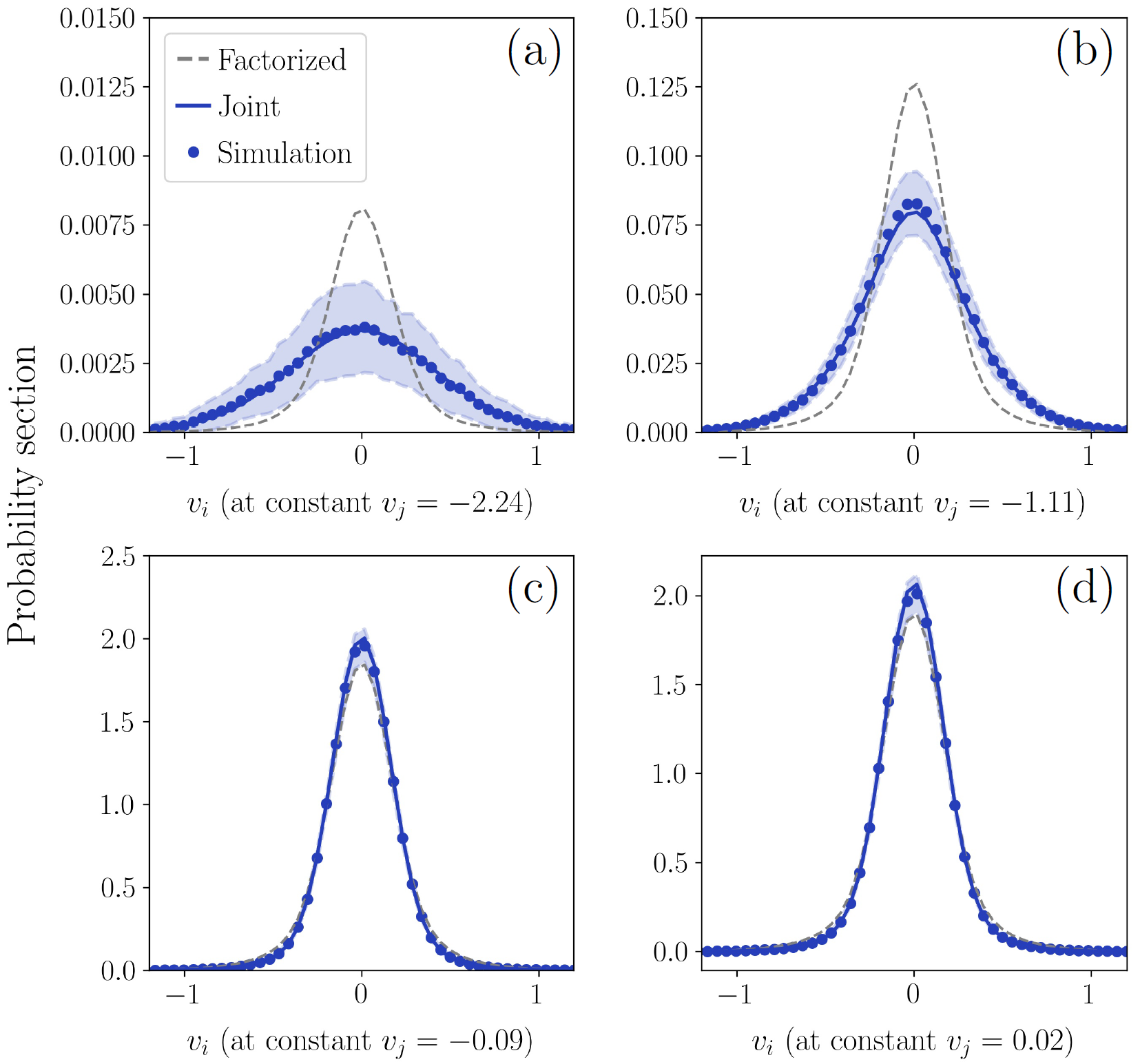}
    \caption{\footnotesize Comparison between the analytical expressions of the joint probability distribution $p(v_i, v_j)$ given by Eq. (\ref{eqn:pstat_joint_si}) and the one obtained from with $10^3$ simulations. We also show the corresponding results for the factorized probabilities. The blue line corresponds to the analytical expression of $p(v_i, v_j)$. The corresponding dots represent the histogram of the distribution obtained from $10^3$ simulations, and the semitransparent filled areas represent one standard deviation from this estimate. Similarly, the gray dashed line represent the analytical expression of $p(v_i)p(v_j)$. (a) Section along the $v_j$ direction for small $v_j$, so that we are looking at the tails of the distribution. Even though the estimate along the tails is noisy, we clearly see that the estimate from the simulations lies along the analytical prediction. (b) As before, but for higher $v_j$. (c-d) As before, but with values of $v_j$ close to zero so we look at the bulk of the distribution near its peak. Even though joint probability and its factorization now are more similar, once more the estimate from the simulation match the analytical expression $p(v_i, v_j)$.}
    \label{fig:probability_data_comparison}
\end{figure}

Let us recall that
\begin{align}
\label{eqn:pstat_joint_si}
    p(v_i, v_j) = \frac{1+ \text{Erf}\left(\frac{\mathcal{D}^*}{\sqrt{\theta \gamma_D}}\right)}{2\pi \mathcal{D}^* \sqrt{\gamma_i\gamma_j}}e^{-\frac{1}{\mathcal{D}^*}\left(\frac{v_i^2}{\gamma_i}+\frac{v_j^2}{\gamma_i}\right)} + \frac{1}{\sqrt{\gamma_i\gamma_j\gamma_D \pi^3\theta}} \int_{\mathcal{D}^*}^\infty \frac{dD}{D} e^{-\frac{1}{D}\left(\frac{v_i^2}{\gamma_i}+\frac{v_j^2}{\gamma_i}\right)}e^{-\frac{D^2}{\theta \gamma_D}}
\end{align}
is the stationary joint probability of the model. In Figure \ref{fig:probability_single}a we check that the stationary distribution of $\mathcal{D}$ is indeed the one presented in the main text and in Figure  \ref{fig:probability_single}b that the stationary distribution of a single $v_i$ corresponds to the analytical expression we derived. If we compare this distribution to a standard distribution of an Ornstein-Uhlenbeck process with a diffusion coefficient equal to the mean $\ev{\mathcal{D}(t)}$ we immediately see that the distribution of our model is considerably more peaked around zero and displays longer tails. Indeed, one expects that due to the fact that $\mathcal{D}^* < \ev{\mathcal{D}(t)}$ the system tends to wander more easily close to zero, especially in the time windows where the diffusion coefficient is constant and equal to $\mathcal{D}^*$. At the same time, the fact that $\mathcal{D}(t)$ can change in time favors the presence of values of $v$ that are larger in absolute value, which is the mechanism at the origin of the bursty behavior seen in the main text.

In Figure \ref{fig:probability_joint_factor} we look instead at the properties of the joint probability distribution $p(v_i, v_j)$. The most natural quantity to compare this distribution with is its factorization $p(v_i)p(v_j)$, which is equivalent to ignoring the feedback effects between $v_i$ and $v_j$ due to the shared extrinsic modulation of $\mathcal{D}(t)$. Since we are setting $\mathcal{D}^* = 0.5$, we expect that these effects are going to be particularly relevant for the dynamics of the model. In particular, in Figure \ref{fig:probability_joint_factor}c-d we see that the most important differences between the two occur in the tails of the two-dimensional distribution, with the joint distribution typically showing dramatically longer tails. This translates to the fact that far-from-zero values of the two variables can occur more easily at the same time. The situation is completely reversed when we increase $\mathcal{D}^*$. In Figure \ref{fig:probability_joint_factor}e-f we see that for $\mathcal{D}^* = 8$ the joint probability distribution and the factorized distribution are almost indistinguishable. Hence, this example shows explicitly that if $\mathcal{D}^*$ is high enough the dependence induced by the extrinsic modulation vanishes.

Let us keep focusing on the case $\mathcal{D}^* = 0.5$ for the time being. In Figure \ref{fig:probability_data_comparison}a-d we compare the one-dimensional sections of the analytical expression of the joint probability distribution with the results of $10^3$ simulations of the model, together with the sections of the factorized distribution. The joint distribution estimated from the simulation  matches particularly well the analytical prediction. Once more, and perhaps more clearly, in Figure \ref{fig:probability_data_comparison}a-b we see the stark difference that emerges along the tails between the joint probability distribution and its factorization. Interestingly, panel (c) and panel (d) show that the situation in the bulk of the distribution is reversed with respect to the tails and now the joint probability distribution is more peaked with respect to is factorization, albeit only slightly. That is, the modulation in the low $\mathcal{D}^*$ regime favors both large values of $v_i$ and $v_j$ and values very close to zero.

Overall, this brief analysis showed us how the bursty behavior that we see for small values of $\mathcal{D}^*$ emerges from the underlying probability distributions, which in turn emerge from the simple marginalization that occurs in Eq. (\ref{eqn:pstat_joint_si}). Similar arguments, albeit impractical, could be carried out for the probability distributions beyond the two-point ones. In a sense, one could argue that the fundamental properties of the model are inherited from the fact that there are some unobserved physical quantities, and these are the quantities that drive the global response of the single variables.

For the specific case of a double Ornstein-Uhlenbeck process we chose, the net effect of the marginalization is the widening of the tails of both the one-point $p(v)$ and the two-point $p(v_i, v_j)$ probability distributions when $\mathcal{D}^*$ is small enough. As we increase $\mathcal{D}^*$, this effect becomes less and less important until it is completely negligible. In this sense, we can effectively think of $\mathcal{D}^*$ as a control parameter that changes the qualitative behavior of the system. Most importantly, the fact that the tails of the joint probability distribution are wider when $\mathcal{D}^*$ is small reflects dynamically in the emergence of a non trivial coordination between the variables, from which in turn power-law avalanches emerge.

\section*{Correlations in the interacting model}
\begin{figure}[h!]
    \centering
    \includegraphics[width=0.95\textwidth]{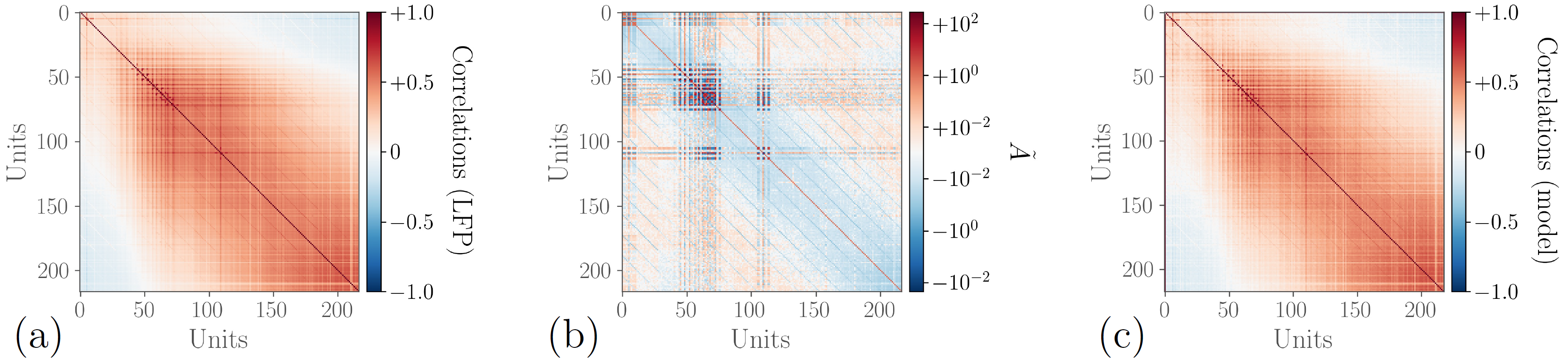}
    \caption{ Comparison between (a) the correlations of the data and (c) the correlations of the interacting model after solving the inverse problem. Panel (b) shows the inferred matrix $\tilde{A}$, which is independent on the parameters of the model.}
    \label{fig:inverse_problem}
\end{figure}

\noindent In Figure \ref{fig:inverse_problem} we compare the correlation matrix obtained from the data, the interaction matrix $\tilde{A}_{ij}$ obtained by solving the inverse problem described in the main text, and the correlation matrix obtained from simulations of the resulting model.

\section*{Avalanches in the extrinsic model, dependence on $\mathcal{D}^*$ and finite-size scaling}
\begin{figure}[h!]
    \centering
    \includegraphics[width = 0.9\textwidth]{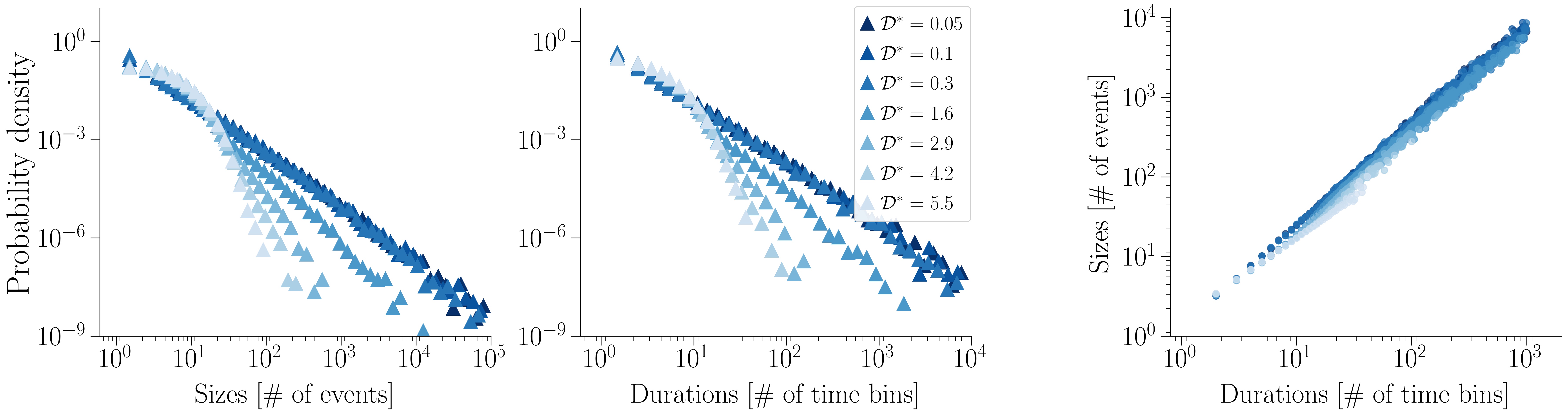}
    \caption{{Avalanches in the extrinsic model at increasing values of $\mathcal{D}^*$.}}
    \label{fig:varyingdstar}
\end{figure}

\noindent 
{Figure \ref{fig:varyingdstar} shows avalanches' distributions in the extrinsic model as $\mathcal{D}^*$ is increased. It is possible to see that, while the exponents converge to a fixed value for $\mathcal{D}^*$ sufficiently low, as  $\mathcal{D}^*$ is increased they gradually increase in absolute value, and the distribution tends to an exponential when $\mathcal{D}^* > 5$. That is, the shift from exponential and power-law avalanches is smooth.}

\begin{figure}[h!]
    \centering
    \includegraphics[width = 0.9\textwidth]{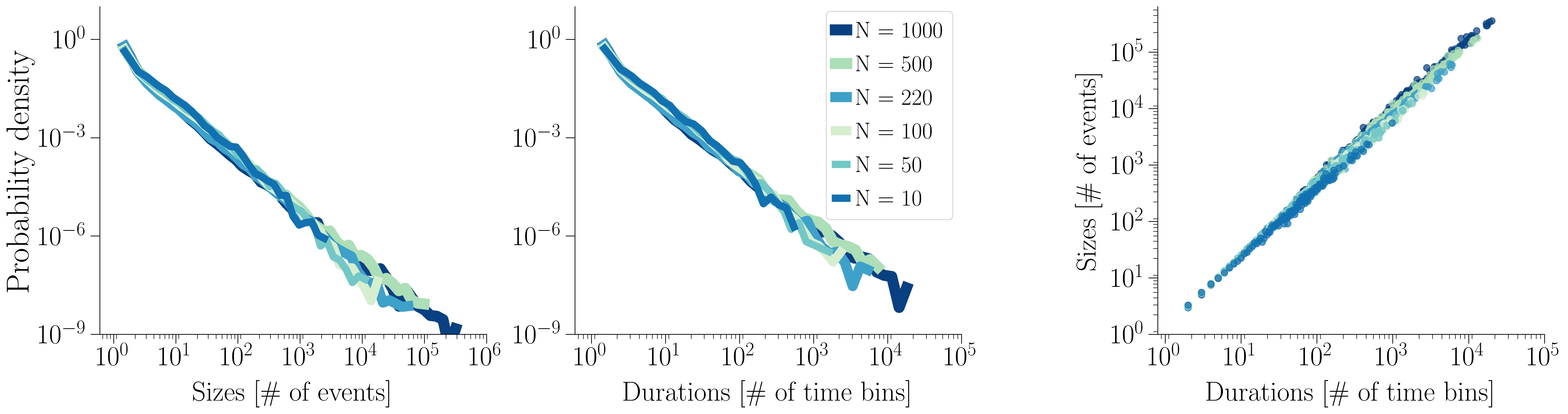}
    \caption{{Finite-size scaling in avalanches found in the extrinsic model.}}
    \label{fig:finitesize}
\end{figure}

{Figure \ref{fig:finitesize} shows instead the finite-size scaling behavior of avalanches in the extrinsic model. The avalanches obtained do not present a clear cutoff. Yet, it is evident that when increasing the number of units the values of the maximum avalanches size and duration increase, while the exponent of the distribution remains the same, displaying a property typical of true power laws. This is perhaps unsurprising, since the units are conditionally independent to begin with. In \cite{Mariani2021b} (SI) we have studied the avalanche finite-size scaling in LFPs from the same set of experiments, and they indeed reproduce the expected finite-size scaling features, as already found in the literature \cite{beggs2003}.}

\begin{figure}[h!]
    \centering
    \includegraphics[width=0.95\textwidth]{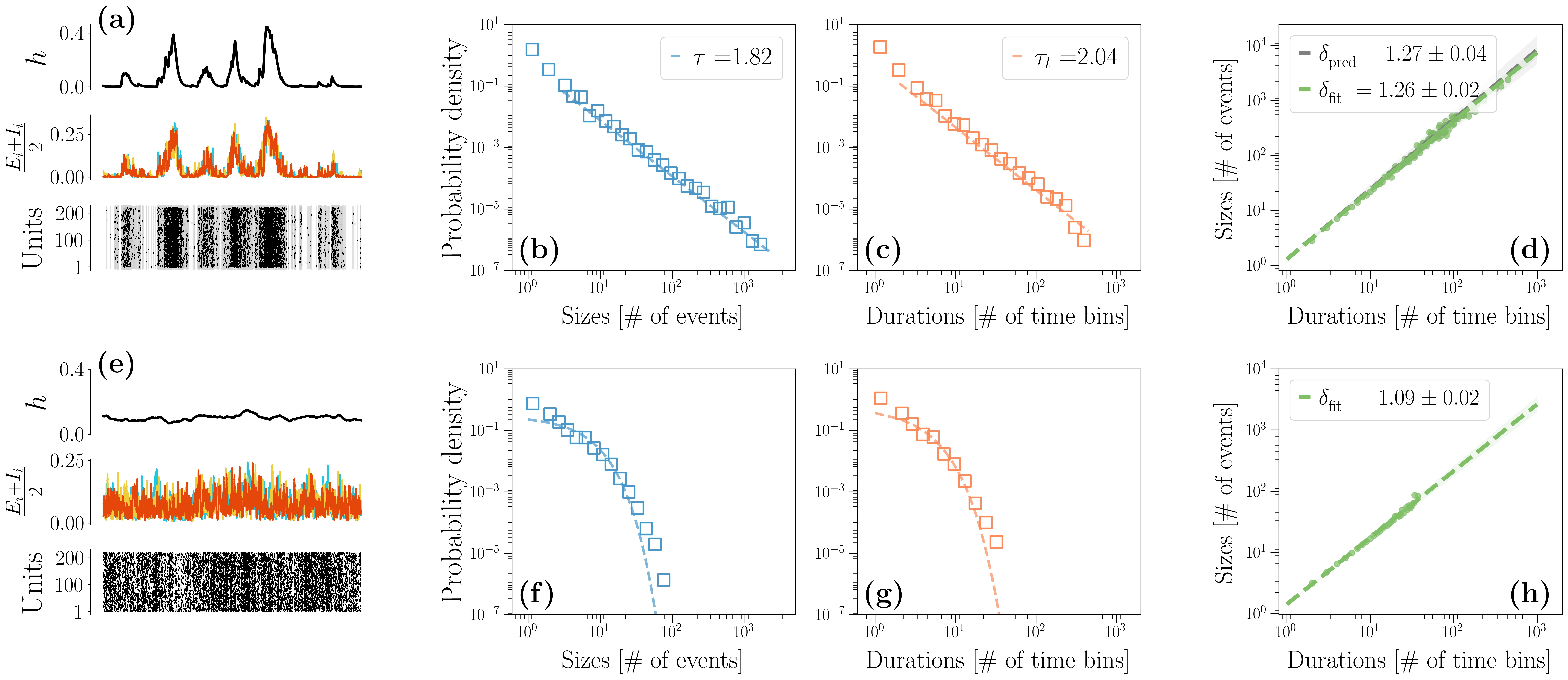}
    \caption{ {Avalanche statistics generated by the Wilson-Cowan units. The ``internal'' Wilson-Cowan units are always in an inhibition dominated phase, i. e. $\omega_{I} = 7$ and $\omega_{E} = 6.8$, and $\alpha = 1$. Their external input $h$ is instead always in a critical state, in particular $\omega^{(h)}_{E} = 50.05$, $\omega^{(h)}_{I} = 49.95$ and $\alpha^{(h)} = 0.1$,  with $h^{(h)} = 10^{-3}$. In panels (a-d) The amplitude of the noise,  is increased to $1.2 \times 10^{-3}$. In panels (e-h) instead the noise is reduced to $5 \times 10^{-5}$. In panels (a) and panel (e) we compare the trajectories of $h$, $\frac{E_i + I_i}{2}$ and the corresponding trains of events in the high (a) and low (e) $\sigma^{(h)}$ regime. (b-d) If $\sigma^{(h)}$ is high avalanches are power-law distributed and the crackling-noise relation is verified. (f-g) Same plots, now in the low $\sigma^{(h)}$ regime. Avalanches are now fitted with an exponential distribution. (h) The average avalanche size as a function of the duration scales with an exponent that, as $\sigma^{(h)}$ decreases, becomes closer to the trivial one $\delta_\mathrm{fit} \approx 1$.}
    }
    
    \label{fig:criticalwilson}
\end{figure}

\section*{A critical stochastic Wilson-Cowan model as external modulation}
\noindent {In Figure \ref{fig:criticalwilson} we analyze some Wilson-Cowan units that receive the same input by another Wilson-Cowan population that is in a critical state. Indeed, in this case we set $\alpha^{(h)} = \omega_0^{(h)} = \omega_E^{(h)} - \omega_I^{(h)} = 0.1$, and we are exactly at the critical point of the stochastic Wilson-Cowan model as shown in \cite{decandia2021}. Here, avalanches in principle will be present for all noise amplitudes (i.e., system sizes). However note that in Figure \ref{fig:criticalwilson} we set $h^{(h)} = 10^{-3}$, so that the dynamics slightly deviates from the critical point defined for $h = 0$. For this reason, avalanches disappear if we reduce the noise to a sufficiently low value (see Figure \ref{fig:criticalwilson}e-h). Notably, the noise amplitude has to be dramatically reduced in order not to see avalanches, at $\sigma^{(h)} = 5 \times 10^{-5}$, corresponding to very large system sizes ($K = L \approx 4 \times 10^7$, with $K$ and $L$ respectively the number of excitatory and inhibitory neurons \cite{Benayoun} corresponding to the neural population).}

\section*{Avalanches statistics across rats}
\begin{figure}[h!]
    \centering
    \includegraphics[width=0.95\textwidth]{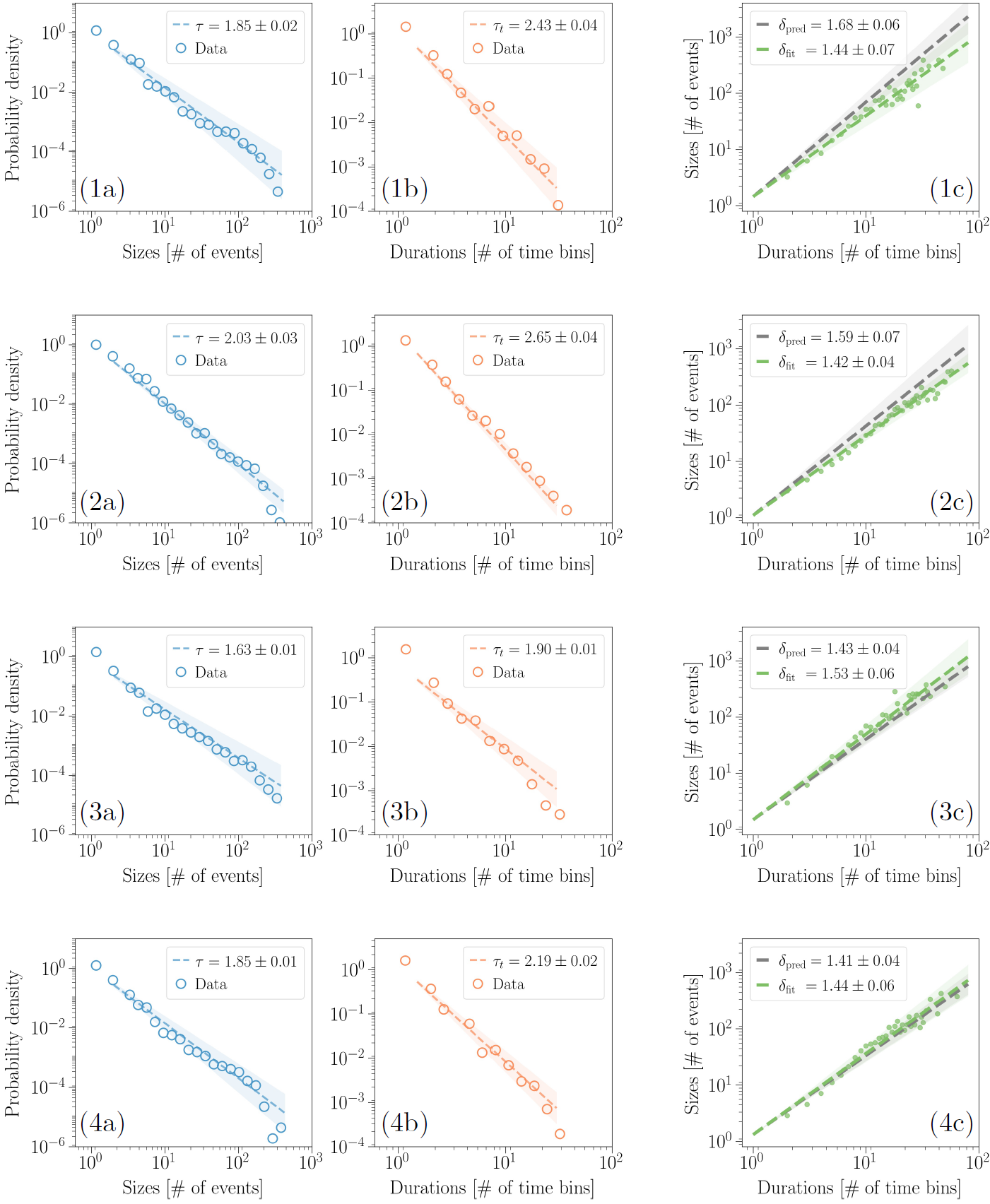}
    \caption{Avalanche statistics in four different rats. (1a-4a) The distribution of the avalanches' sizes is consistently a power-law, with an exponent that slightly depends on the single rat. (1b-4b) The avalanche durations are once more power-law distributed in all rats with some variability in the exponents, even though the range accessible with the experimental setup only covers two decades. (1c-4c) The crackling-noise relation, however, is consistently satisfied in each rat.}
    \label{fig:resting_avalanches}
\end{figure}

\noindent In Figure \ref{fig:resting_avalanches} we report the avalanches statistics from the other rats that we analyzed. The avalanches statistics is computed by considering all available the $20$ trials of basal activity for each rat, that are $7.22 \mathrm{s}$ long. Inter-rat variability is present with respect to avalanche exponents, and is expected as found in previous experiments \cite{Fontenele2019, Shew2015, bansal2020}. Moreover, a theoretical explanation for the difference in avalanche exponents has been also recently proposed as signature of quasi-criticality \cite{Fosque2021}. However, the fundamental point is that the crackling-noise relation is always verified compatibly with the experimental errors, a feature that is usually considered a hallmark of criticality \cite{Fontenele2019}.

%\bibliography{paper_bib}
\end{document}